\documentclass[twocolumn,preprintnumbers,superscriptaddress]{revtex4-2}
\usepackage{amsmath}
\usepackage{amssymb}
\usepackage{mathrsfs}
\usepackage{graphicx}
\usepackage{bm}
\usepackage{epstopdf}
\usepackage[colorlinks=true,citecolor=blue,linkcolor=blue,urlcolor=blue]{hyperref}
\usepackage{color}
\begin{document}
\title{Magneto-optical conductivity of monolayer transition metal dichalcogenides
in the presence of proximity-induced exchange interaction and external electrical field}

\author{Y. Li}
\affiliation{School of Physics and Astronomy and Yunnan Key Laboratory of Quantum
Information, Yunnan University, Kunming 650091, China}

\author{Y. M. Xiao}\email{yiming.xiao@ynu.edu.cn}
\affiliation{School of Physics and Astronomy and Yunnan Key Laboratory of Quantum
Information, Yunnan University, Kunming 650091, China}

\author{W. Xu}\email{wenxu\_issp@aliyun.com}
\affiliation{School of Physics and Astronomy and Yunnan Key Laboratory of Quantum
Information, Yunnan University, Kunming 650091, China}
\affiliation{Key Laboratory of Materials Physics, Institute of Solid State
Physics, HFIPS, Chinese Academy of Sciences, Hefei 230031, China}
\affiliation{Micro Optical Instruments Inc., Shenzhen 518118, China}

\author{L. Ding}
\affiliation{School of Physics and Astronomy and Yunnan Key Laboratory of Quantum
Information, Yunnan University, Kunming 650091, China}

\author{M. V. Milo\v{s}evi\'{c}}
\affiliation{Department of Physics, University of Antwerp,
Groenenborgerlaan 171, B-2020 Antwerpen, Belgium}

\author{F. M. Peeters}
\affiliation{Department of Physics, University of Antwerp, Groenenborgerlaan 171,
B-2020 Antwerpen, Belgium}
\affiliation{Micro Optical Instruments Inc., Shenzhen 518118, China}

\date{\today}

\begin{abstract}
We theoretically investigate the magneto-optical (MO) properties of monolayer (ML)
transition metal dichalcogenides (TMDs) in the presence of external electrical
and quantizing magnetic fields and of the proximity-induced exchange interaction.
The corresponding Landau Level (LL) structure is studied by solving the Schr\"odinger
equation and the spin polarization in ML-TMDs under the action of the magnetic field
is evaluated.The impact of trigonal warping on LLs and MO absorption is examined.
Furthermore, the longitudinal MO conductivity is calculated through the dynamical
dielectric function under the standard random-phase approximation (RPA) with the Kubo formula.
We take ML-MoS$_2$ as an example to examine the effects of proximity-induced exchange
interaction, external electrical and magnetic fields on the MO conductivity
induced via intra- and interband electronic transitions among the LLs. For intraband
electronic transitions within the conduction or valence bands, we can observe two
absorption peaks in terahertz (THz) frequency range. While the interband electronic
transitions between conduction and valence LLs show a series of absorption peaks in
the visible range. We find that the proximity-induced exchange interaction, the carrier
density, the strengths of the external electrical and magnetic fields can effectively
modulate the positions of the absorption peaks and the shapes of the MO absorption
spectra. The results obtained from this study can benefit to an in-depth understanding
of the MO properties of ML-TMDs which can be potentially applied for magneto-optic,
spintronic and valleytronic devices working in visible to THz frequency bandwidths.
\end{abstract}

\maketitle

\section{Introduction}
Since the discovery of graphene in 2004 \cite{Novoselov04}, the investigation
of atomically thin or two-dimensional (2D) electronic materials has become an
important and fast-growing field of research field in condensed matter physics
and electronics. In recent years, monolayer (ML) transition metal
dichalcogenides (TMDs) have attracted a great attention due to their excellent
and unique physical properties such as the spintronic and valleytronic features
for potential applications in, e.g., information technology. From a viewpoint
of condensed matter physics, ML-TMDs is in the form of $X$-$M$-$X$ with two layers of
chalcogen atoms ($X$=S, Se, Te) in two hexagonal planes separated by a plane of
metal atoms ($M$=Mo, W, Nb, Ta, Ti) \cite{Wang12}. ML-TMDs has the direct band
gaps around the $K$- and $K'$-point, whereas the multi-layer TMDs are normally
with indirect gaps \cite{Mak10,Splendiani10}. Due to the presence of the intrinsic
spin and the Rashba spin-orbit coupling (SOC), the spin degeneracy in ML-TMDs
is often lifted and the spin orientations in energy states in different valleys
are just opposite. The corresponding spin polarization can be observed experimentally
by using, e.g., optical measurements with circularly polarized radiation
field \cite{Mak14,Xiao17}. Moreover, it has been found that the valley degeneracy
in ML-TMDs can be lifted via, e.g., introducing the exchange interaction induced by,
e.g., the proximity effect in the presence of ferromagnetic substrate \cite{Qi15,Cortes22,Zhao20}.
Therefore the spin-orbit and spin-valley couplings in ML-TMDs can contribute to
the unique spin and valley degree of freedoms. As a result, ML-TMDs can be applied
as spin and valley filter \cite{Zambrano21}, gas-sensor \cite{Li12s}, ultra-sensitive
photodetectors \cite{Yin12}, and field-effect transistors (FETs) with on/off
ratios of the order of 10$^8$ and mobility larger than 200 cm$^2$V$^{-1}$s$^{-1} $\cite{Radisavljevic13}.
These interesting characteristics suggest that ML-TMDs based 2D electronic materials are
promising candidates for next generation of high performance devices in
nano-electronics \cite{Radisavljevic11}, optoelectronics \cite{Lopez-Sanchez13,Xiao16},
twistronics \cite{Ciarrocchi22}, spintronics and valleytronics \cite{Zeng12N,Mak12N,Cao12,Mak14}.\par

The investigation of the electronic and optoelectronic properties of two
dimensional electron gas (2DEG) in the presence of an external magnetic
field is an intriguing field of research \cite{Ando82,Giuliani05,CastroNeto09,Goerbig11,Tabert13}.
More specifically, the magneto-optical (MO) properties of ML-TMDs have attracted
lots of attentions in recent years \cite{Rose13,Chu14,Mitioglu15,Catrina19}.
When a quantizing magnetic field is applied perpendicularly to the 2D plane of
a ML-TMDs, the Landau quantization can lead to the Landau levels (LLs) in the
energy spectrum and the electronic transitions among these LLs can result
in many interesting physical phenomena such as the integer quantum Hall
effect \cite{Tahir16,2Tahir16}, fractional quantum Hall effect \cite{Dean20},
Shubnikov-de Haas (SdH) oscillations \cite{Pisoni18,Smolenski19}, photo-induced
anomalous Hall effect \cite{Nguyen21}, etc. As we know, the MO conductivity is
a key physical quantity which is directly related to the MO absorption measured
experimentally via transmission and reflection. The theoretical calculation of
MO conductivity of ML-TMDs was carried out by taking into account of spin and
valley polarizations, electron-phonon coupling, spin and valley Zeeman effects,
and nonlinear effect \cite{Li12,Tahir16,2Tahir16,Nguyen17,Hien20}. It has been
shown that the spin and valley dependent MO absorption induced by the electronic
transitions among LLs can occur in visible to terahertz (THz) bandwidth. These
unique and interesting MO properties of ML-TMDs can be utilized to design and apply
for the advanced MO materials and devices for applications from visible to THz bandwidths.

Moreover, due to the unique feature of light-valley coupling \cite{Xiao12,Mak18Na},
ML-TMDs can exhibit the novel transport effects such as valley-selective circular
dichroism \cite{Cao12}, valley Hall effect (VHE) \cite{Mak14}, and Faraday rotation \cite{Da17}.
The enhancement of valley splitting and a large valley polarization of ML-TMDs can
be achieved through placing it on ferromagnetic substrate such as EuO via the
proximity-induced exchange interaction \cite{Qi15,Cortes22}. The techniques for
tuning the spin and/or valley degree of freedoms of ML-TMDs are prominent research
fields for the time being. It should be noted that the proximity-induced exchange
interaction can lift the valley degeneracy in electronic energy spectrum in ML-TMDs.
Apparently, the optical \cite{Zhao20} and MO properties of ML-TMDs can be affected
by the lifting of the valley degeneracy. Thus, it is interesting and significant
to investigate the MO properties of ML-TMDs with the breaking of the spin and valley
degeneracies by the proximity-induced exchange interaction. This becomes the prime
motivation of this study. In this work, we study theoretically the Landau level structure
and the longitudinal MO conductivity of both $n$-type and $p$-type ML-TMDs in the
presence of proximity-induced exchange interaction which can be induced by the substrate
together with external electrical and magnetic fields. The effects of spin and valley
polarizations, magnetic field strength, electrical field strength, proximity-induced
exchange interaction, and carrier density on the MO conductivity of ML-MoS$_2$ are
examined theoretically.

The present paper is organized as follows. In Sec. \ref{sec:theoretical approach},
we evaluate the LL structure and the longitudinal magneto-optical conductivity of
both $n$-type and $p$-type ML-TMDs with the framework of dynamical dielectric function
on the basis of the random-phase approximation (RPA). In Sec. \ref{sec:results},
we analyze and discuss the influence of spin and valley polarizations on MO conductivity
induced by electronic transitions within the conduction/valence LLs and between the
conduction and valence LLs in ML-TMDs such as ML-MoS$_2$. The main conclusions obtained
from this study are summarized in Sec. \ref{sec:conclusions}.

\section{Theoretical approach}
\label{sec:theoretical approach}
\subsection{Landau level structure}
We consider a ML-TMDs laid in the $x$-$y$ plane on a ferromagnetic substrate
which can lead to an
exchange interaction in the presence of an external perpendicular electrical
field $\mathbf{E}=(0,0,E_z)$. The Hamiltonian for an electron in ML-TMDs at $K$ ($K'$)
valley can be written as \cite{Hien20,Qi15,Zhao20,Ochoa13}
\begin{align}\label{hamil}
H^\zeta(\mathbf{k})=&\hbar v_\mathrm{F}(\zeta k_x\hat{\sigma}_x+k_y\hat{\sigma}_y)
+(\Delta+edE_z)\hat{\sigma}_z\notag \\
&+\zeta s(\lambda_c\hat{\sigma}_++\lambda_v\hat{\sigma}_-)
-s(B_c\hat{\sigma}_++B_v\hat{\sigma}_-),
\end{align}
where $\mathbf{k}=(k_x,k_y)$ is the carrier wave vector or momentum operator,
$v_\mathrm{F}$ is the Fermi velocity, $\zeta=\pm1$ is the index for
$K$ ($K^\prime$) valley, $s=\pm 1$ is a good quantum number for spin up/down
states (implying that the spin-flip transitions are prohibited in the system), $\hat{\sigma}_i$
and $\hat{s}_i$ are the Pauli matrices
of the sublattice pseudo-spin and the real spin respectively, $\hat{\sigma}_0 $
is an unit $2\times2 $ identity matrix,
$\hat{\sigma}_\pm=(\hat{\sigma}_0\pm\hat{\sigma}_z)/2$, $\Delta$ is the band
gap parameter of the ML-TMDs, 2$d$ is the vertical distance
between the $M$ (e.g., Mo) and $X$ (e.g., S) sublattices in ML-TMDs, and
$E_z$ is the strength of the applied perpendicular electrical field. We note
that the term $edE_z\hat{\sigma}_z$ is induced on the basis of the fact that
the conduction and valence band edges near $K$ and $K^{'}$ points
in ML-TMDs are attributed mainly from $d_{z^2}$, $d_{xy}$ and $d_{x^{2}-y^{2}}$
orbits of the metal atoms. The thickness of the ML-TMD is $4d$.
In Eq. \eqref{hamil}, $\lambda_c$ and $\lambda_v$ are the intrinsic spin-orbit coupling (SOC) parameters
in conduction and valence band. $B_c$ and $B_v$ are the experienced effective
Zeeman fields originated from the proximity-induced exchange interaction in
conduction and valence bands, respectively. It should be noted
that the parameters $B_c$ and $B_v$ should depend on the band alignment between the
TMDs layer and the ferromagnetic substrate and on the surface termination of the
substrate. They can be obtained through, e.g., ab initio calculations \cite{Qi15,Norden19}.
In the present study, we take $B_c$ and $B_v$ as the input parameters in the following
calculations. Furthermore, the band gap parameter $\Delta$ is always present in ML-TMDs. As we
can see from the second term on the righthand side of Eq. \eqref{hamil}, the actual
band gap in ML-TMDs in the presence of pseudo-spin induced by Pauli matrix $\sigma_z$
can be affected by the presence of the perpendicular electric field $E_z$.
This Hamiltonian can also be written in the form of a matrix
\begin{align}
H^\zeta=\left(\begin{array}{cc}
\Delta_{\zeta s}&\zeta \hbar v_\mathrm{F}k_{-\zeta}\\
\zeta \hbar v_\mathrm{F}k_{\zeta}&-\Delta_{\zeta s}
\end{array}\right)+O_{\zeta s},
\end{align}
where $k_{\pm}=k_x\pm ik_y=ke^{\pm i\theta}$, $\theta$ is the angle between
$\mathbf{k}$ and the $x$ axis, and
$O_{\zeta s}=\zeta s(\lambda_c+\lambda_v)/2-s(B_c+B_v)/2$. The corresponding
Schr\"odinger equation can be solved analytically and the eigenvalues are obtained as
\begin{equation}
\varepsilon_{\beta}^{\zeta s}(\mathbf{k})=
\beta \sqrt{\hbar^2 v^2_\mathrm{F}k^2+\Delta_{\zeta s}^2}+O_{\zeta s},\label{bandstructure}
\end{equation}
where $\Delta_{\zeta s}=\Delta+\zeta s(\lambda_c-\lambda_v)/2-s(B_c-B_v)/2+edE_z$
with $\beta=\pm 1$ refers to conduction and valence band, respectively.

In the presence of a perpendicular magnetic field $\mathbf{B}=(0,0,B)$ applied
in the Faraday geometry on a ML-TMDs layer, the Hamiltonian for an electron in
ML-TMDs now becomes \cite{Hien20,Zhao20,Qi15,Tahir16}
\begin{align}
H^\zeta=&v_\mathrm{F}(\zeta\Pi_x\hat{\sigma}_x+\Pi_y\hat{\sigma}_y)
+(\Delta+edE_z)\hat{\sigma}_z\notag \\
&+\zeta s(\lambda_c\hat{\sigma}_++\lambda_v\hat{\sigma}_-)
-s(B_c\hat{\sigma}_++B_v\hat{\sigma}_-)\notag \\
&+sM_s-\zeta M_v.\label{hamilmag}
\end{align}
Here, $\mathbf{\Pi} =\mathbf{p}+ e\mathbf{A}$ is the canonical momentum
with a vector potential $\mathbf{A}$, $M_j$ = g$_j\mu_BB/2$, the Bohr magneton
$\mu_B=e\hbar/2m_e$ with $m_e$ being the rest electron mass,
$j=s,v$ stands for the spin or valley Zeeman field,
g$_s$ = g$_e$ + g$_s^\prime$, g$_e=2$ is the g-factor for an free-electron,
and g$_s^\prime$ is out-of-plane effective spin g-factor due to the presence
of the SOC \cite{Tahir16,Kormanyos14}. We have taken the Landau gauge
$\mathbf{A}=(0,Bx,0)$ for the vector potential of the magnetic field for the
theoretical calculation in this study.

After defining the ladder operators with the magnetic field,
$a_{\pm}=l(\Pi_x\pm i\Pi_y)/(\sqrt{2}\hbar)$ which acts on the wave functions
of the harmonic oscillator and $l=\sqrt{\hbar/eB}$ is the magnetic length, the
Hamiltonian given by Eq. \eqref{hamilmag} can be written as
\begin{align}
H^\zeta=&\left(\begin{array}{cc}
\Delta_{\zeta s}&\zeta\hbar\omega_ca_\mp\\
\zeta\hbar\omega_ca_\pm&-\Delta_{\zeta s}
\end{array}\right)+O_{\zeta s}+sM_s-\zeta M_v,
\end{align}
where $\omega_c=\sqrt{2}\upsilon_\mathrm{F}/l$
is the cyclotron frequency for a ML-TMDs system. After diagonalizing this Hamiltonian,
we obtain the electronic energy for a state $\alpha=(\zeta,s,\beta,n,k_y)$ as
\begin{align}
E_\alpha=E_{\beta n}^{\zeta s}=\beta h_{n}^{\zeta s}+O_{\zeta s}
+sM_s-\zeta M_v, \label{egvalue}
\end{align}
where $h_{n}^{\zeta s}=[n\hbar^2\omega^2_c+\Delta_{\zeta s}^2]^{1/2}$ and $n$ is
the LL index. The corresponding eigenfunctions $\psi_{\beta n k_y}^{\zeta s}$ for the $K$
and $K^\prime$ valleys are given respectively by \cite{Tahir16}
\begin{align}
\psi_{\beta n k_y}^{+ s}=\frac{e^{ik_yy}}{\sqrt{L_y}} \left(\begin{array}{c}
C_{\beta n}^{+ s}\phi_{n-1}(\overline{x})\\
D_{\beta n}^{+ s}\phi_{n}(\overline{x})\end{array}\right),
\end{align}
and
\begin{align}
\psi_{\beta n k_y}^{- s}=\frac{e^{ik_yy}}{\sqrt{L_y}} \left(\begin{array}{c}
-C_{\beta n}^{- s}\phi_{n}(\overline{x})\\
D_{\beta n}^{- s}\phi_{n-1}(\overline{x})\end{array}\right).
\end{align}
Here, the normalization constants are
$C_{\beta n}^{\zeta s}=[1/2+\beta\Delta_{\zeta s}/(2h_{n}^{\zeta s})]^{1/2}$
and $D_{\beta n}^{\zeta s}=[1/2-\beta\Delta_{\zeta s}/(2h_{n}^{\zeta s})]^{1/2}$,
$k_y$ is the electron wave vector along the $y$ direction, $\overline{x}=x-x_0$ with $x_0=l^2k_y$, and $\phi_n(x)=e^{-x^2/2}H_n(x)/\sqrt{2^nn!\sqrt{\pi}l}$ with $H_n(x)$ being
the Hermite polynomials. Specially, the eigenfunctions for $n=0$ at the $K$
and $K^\prime$ valleys are given respectively as
\begin{align}
\psi_{\beta 0 k_y}^{+ s}=\frac{e^{ik_yy}}{\sqrt{L_y}}\left(\begin{array}{c}
0\\
\phi_{0}(\overline{x})\end{array}\right) \ {\rm and} \
\psi_{\beta 0 k_y}^{- s}=\frac{e^{ik_yy}}{\sqrt{L_y}}\left(\begin{array}{c}
\phi_{0}(\overline{x})\\
0\end{array}\right).
\end{align}

\subsection{Effect of trigonal warping}
In the presence of an external magnetic field, the trigonal warping can lead to
higher-order corrections to electronic band structure in ML-TMD. This effect can be accounted via the
term \cite{Rose13,Kormanyos13}
\begin{align}
H^\zeta_{tw}=-\frac{2\gamma}{l^2}&\left(\begin{array}{cc}
0&a_\pm^2\\
a_\mp^2&0
\end{array}\right),
\end{align}
where the relevant parameter $\gamma=-1.02$ eV{\AA}$^2$ is associated
with the trigonal warping and can be obtained from fitting of the energy spectra to tight-binding
or ab initio calculations \cite{Rose13,Kormanyos13}. By using the perturbation
theory, we can take the trigonal warping as a second-order correction
relative to the leading LLs. For the $K$ and $K^\prime$ valleys, the second-order
corrections $E_\alpha^{(2)}$ are given respectively as
\begin{align}
E_\alpha^{(2)}=&\frac{4\gamma^2}{l^4}[\mathbb{A}_{n\beta}^{+ s}C_{\beta n+3}^{+ s}D_{\beta n}^{+ s}\sqrt{(n+1)(n+2)}\nonumber\\
&+\mathbb{B}_{n\beta}^{+ s}C_{\beta n}^{+ s}D_{\beta n-3}^{+ s}\sqrt{(n-1)(n-2)}],
\end{align}
with
$\mathbb{A}_{n\beta}^{+ s}=C_{\beta n+3}^{+ s}D_{\beta n}^{+ s}\sqrt{(n+1)(n+2)}/(E_{\beta n}^{+ s}-E_{\beta n+3}^{+ s})$ and $\mathbb{B}_{n\beta}^{+ s}=C_{\beta n}^{+ s}D_{\beta n-3}^{+ s}\sqrt{(n-1)(n-2)}/(E_{\beta n}^{+ s}-E_{\beta n-3}^{+ s})$,
and
\begin{align}
E_\alpha^{(2)}=&\frac{4\gamma^2}{l^4}[\mathbb{A}_{n\beta}^{- s}C_{\beta n}^{- s}D_{\beta n+3}^{- s}\sqrt{(n+1)(n+2)}\nonumber\\
&+\mathbb{B}_{n\beta}^{- s}C_{\beta n-3}^{- s}D_{\beta n}^{- s}\sqrt{(n-1)(n-2)}],
\end{align}
with $\mathbb{A}_{n\beta}^{- s}=C_{\beta n}^{- s}D_{\beta n+3}^{- s}\sqrt{(n+1)(n+2)}/(E_{\beta n+3}^{- s}-E_{\beta n}^{- s})$ and $\mathbb{B}_{n\beta}^{- s}=C_{\beta n-3}^{- s}D_{\beta n}^{- s}\sqrt{(n-1)(n-2)}/(E_{\beta n-3}^{- s}-E_{\beta n}^{- s})$.
The modified eigenstate $\psi_{\alpha}^{(1)}$ with the first order at $K$
and $K^\prime$ valleys are given respectively by
\begin{align}
\psi_{\alpha}^{(1)}=&-\frac{2\gamma }{l^2}(\mathbb{A}_{n\beta}^{+ s}\psi_{\beta, n+3, k_y}^{+ s}+\mathbb{B}_{n\beta}^{+ s}\psi_{\beta, n-3, k_y}^{+ s}),
\end{align}
and
\begin{align}
\psi_{\alpha}^{(1)}=&-\frac{2\gamma}{l^2}(\mathbb{A}_{n\beta}^{- s}\psi_{\beta, n+3, k_y}^{- s}+\mathbb{B}_{n\beta}^{- s}\psi_{\beta, n-3, k_y}^{- s}).
\end{align}
The total eigenfunction is
\begin{align}
\psi'_{\alpha}=\mathbb{F}(\psi_{\beta n k_y}^{\zeta s}+\psi_{\alpha}^{(1)}),
\end{align}
with $\mathbb{F}=l^2/\{l^4+4\gamma^2[(\mathbb{A}_{n\beta}^{\zeta s})^2
+(\mathbb{B}_{n\beta}^{\zeta s})^2]\}^{1/2}$ being the normalization factor.
The effect of trigonal warping can result in additional allowed transition
channels $n$$\rightarrow$ $n$ $\pm$ 2 and $n$$\rightarrow$ $n\pm 4$.
However, it has been shown that the spectral weights of the magneto-optical
absorption spectra induced by trigonal warping is significantly lower than those induced by
 the usual dipolar transitions \cite{Rose13}.

\subsection{Density-of-states and spin-polarization}
The density-of-states (DOS) per unit area can be obtained through the imaginary
part of the Green's function via
\begin{align}
D(E)=\frac{1}{S_0}\sum_{\zeta,s,n,k_y}\mathrm{Im}\ G_\alpha(E),\label{Dos}
\end{align}
where $G_\alpha(E)=(E-E_\alpha+i\delta)^{-1}$ is the retarded Green's function, $E$ is
the carrier energy, and $S_0$ the area of the sample system. With the non-degenerate
statistics in our system, the summation over $k_y$ can be replaced
by an integration \cite{Tahir16,2Tahir16}
\begin{align}
\sum_{k_y}\rightarrow\frac{L_y}{2\pi}\int_{-k_0}^{k_0}dk_y=\frac{S_0}{D_0},
\end{align}
where $k_0= L_x/(2l)^2$ and $D_0 = 2\pi l^2$. Therefor, Eq. \eqref{Dos} can also
be written as
\begin{align}
D(E)=\frac{1}{D_0}\sum_{\zeta, s, n} \delta(E-E_{\beta n}^{\zeta s}).
\end{align}

At a finite temperature, the electron (hole) density $n_{\beta}$ ($\beta= +$ for electrons
in conduction band and $\beta = -$ for holes in valence band) satisfies the electron number
conservation law \cite{Tahir16,Xiao17}
\begin{align}
n_\beta&=\frac{1}{D_0}\sum_{\zeta,s,n}[\delta_{\beta,-1}+\beta f(E_{\beta n}^{\zeta s})],
\end{align}
where $f(x)=[1+{\rm exp}((x-\mu_\beta)/(k_BT))]^{-1}$
is Fermi-Dirac function with $\mu_\beta$ being the chemical
potential (or Fermi energy $E_\mathrm{F}$ at zero temperature) for electrons in $n$-type or
holes in $p$-type sample. The spin polarization of the system is given by
\begin{align}
\mathcal{S}=\frac{n_\beta^{++}+n_\beta^{-+}-n_\beta^{+-}-n_\beta^{--}}{n_\beta},
\end{align}
where $n_\beta=n_\beta^{++}+n_\beta^{-+}+n_\beta^{+-}+n_\beta^{--}$, and $n_\beta^{\zeta s}=
\sum_{n}[\delta_{\beta,-1}+\beta f(E_{\beta n}^{\zeta s})]/D_0$.

\subsection{Dynamical RPA dielectric function and magneto-optical conductivity}
In this study, we study the MO properties of ML-TMDs in the presence of
quantizing magnetic field under the standard random phase approximation (RPA). The PRA
approach has been successfully employed in studying the collective excitations
in 2DEG systems in the presence or absence of an external magnetic
field \cite{Hwang07,tahir08,Roldan11,Goerbig11,Xu03,Kotov12}. The dynamical RPA
dielectric function for ML-TMDs can be written as
\begin{equation}
\epsilon(q,\omega)=1-v_c(q)\Pi(q,\omega),
\end{equation}
where $v_c(q)=2\pi e^2/\kappa q $ is the 2D Fourier transform of the Coulomb
potential, $\kappa$ the effective background dielectric constant of ML-TMDs,
and ${\bf q}=(q_x,q_y)$ is the change of the electron wavevector during an
electron-electron (e-e) scattering event. The non-interacting
density-density ($d$-$d$) correlation function is given by \cite{Tahir15}
\begin{align}
\Pi(q,\omega)=&\frac{1}{S_0}\sum_{\zeta, s}\sum_{\beta n,\beta' n'}\sum_{k_y,k'_y}
\mathbb{C}^{\zeta s}_{\beta nk_y,\beta' n'k'_y}(q)\nonumber\\
&\times\frac{f(E_{\beta n}^{\zeta s})-f(E_{\beta' n'}^{\zeta s})}{E_{\beta n}^{\zeta s}
-E_{\beta' n'}^{\zeta s}+\hbar \omega+i\delta},
\end{align}
where the form factor for e-e interaction is
$$\mathbb{C}^{\zeta s}_{\beta nk_y,\beta' n'k'_y}(q)
=|\langle\alpha'|e^{-i\mathbf{q}\cdot\mathbf{r}}|\alpha\rangle|^2
=V^{\zeta s}_{\beta n,\beta' n'}(u)\delta_{k_y^\prime,k_y-q_y},$$ and
\begin{align}
V^{\zeta s}_{\beta n,\beta' n'}(u)=&[C_{\beta', n'}^{\zeta s}
C_{\beta, n}^{\zeta s}F_{n-1,n'-1}(u)\nonumber\\
&+D_{\beta',n'}^{\zeta s}D_{\beta,n}^{\zeta s}F_{n,n'}(u)]^2,
\end{align}
with $u = l^2q^2/2$. For $n\leq n'$, we have
$[F_{n,n'}(u)]^2=[n!/n'!]e^{-u}u^{n'-n}[L_n^{n'-n}(u)]^2$ and
$L_N^J(x)$ is the Laguerre polynomial.
For the case of $n>n'$, it has the same expression but with that $n$ and
$n'$ are interchanged.
We expand $V^{\zeta s}_{\beta n,\beta' n'}(u)$ to the lowest order ($n'-n=\pm1$) with
\begin{align}\label{eq8}
V^{\zeta s}_{\beta n,\beta'n+1}(u)\rightarrow &
[C_{\beta' n+1}^{\zeta s}C_{\beta n}^{\zeta s}]^2 nu
+[D_{\beta'n+1}^{\zeta s}D_{\beta n}^{\zeta s}]^2(n+1)u \notag\\
&+2C_{\beta' n+1}^{\zeta s}C_{\beta n}^{\zeta s}
D_{\beta' n+1}^{\zeta s}D_{\beta n}^{\zeta s}[n(n+1)]^{1/2}u,
\end{align}
and
\begin{align}\label{eq9}
V^{\zeta s}_{\beta n,\beta'n-1}(u)\rightarrow
&[C_{\beta'n-1}^{\zeta s}C_{\beta n}^{\zeta s}]^2 (n-1)u
+[D_{\beta'n-1}^{\zeta s}D_{\beta n}^{\zeta s}]^2 nu\notag \\
&+2C_{\beta'n-1}^{\zeta s}C_{\beta n}^{\zeta s}
D_{\beta n-1}^{\zeta s}D_{\beta n}^{\zeta s}[n(n-1)]^{1/2}u,
\end{align}
respectively. The real part of the longitudinal MO conductivity for a 2D electronic system can be
derived from the Kubo
formula \cite{Yang07}, which reads
\begin{align}
\sigma_{xx}(\omega)=&\lim_{q\rightarrow 0}
\frac{\kappa\omega}{2\pi q}\mathrm{Im}\ \epsilon(\omega,q)=
-\lim_{q\rightarrow 0}\frac{e^2\omega}{q^2}\mathrm{Im}\ \Pi(q,\omega).
\end{align}
The real and imaginary parts of the ($d$-$d$) correlation function can be
obtained from the Dirac identity
\begin{equation}
\mathrm{lim}_{\eta\rightarrow0^+}\frac{1}{x\pm i\eta}
=\mathcal{P}(\frac{1}{x})\mp i\pi\delta(x),
\end{equation}
where $\mathcal{P}$ denotes the principal value of $1/x$.
Therefore the longitudinal MO conductivity is obtained
as
\begin{align}\label{eq22}
\sigma_{xx}(\omega)=&\lim_{q\rightarrow 0}\frac{\pi\sigma_{0}\hbar\omega}{2u}
\sum_{\zeta,s,n,\beta,\beta'}\delta_{n',n\pm 1}[f(E_{\beta n}^{\zeta s})
-f(E_{\beta' n'}^{\zeta s})]\notag \\
&\times V^{\zeta s}_{\beta n,\beta'n'}\delta(E_{\beta n}^{\zeta s}
-E_{\beta' n'}^{\zeta s}+\hbar \omega),
\end{align}
where $n'=n\pm1$ is given by the selective rules for electronic transitions
and $\sigma_{0}=e^2/h$ is the quantum conductivity.

In the present study, we mainly focus on the
magneto-optical response of ML-TMDs by looking into the real part of the longitudinal MO conductivity $\sigma_{xx}(\omega)$. From the obtained RPA dielectric function, we can also evaluate the imaginary part of the longitudinal MO conductivity.  It is known that the real part of the optical conductivity corresponds to the
energy-consuming process or optical absorption, whereas the imaginary part of it describes the energy exchanging process between
electrons and the radiation field. It should be noted that the dynamical dielectric function and the corresponding MO conductivity obtained in this study are the consequence of the density-density correlation induced by e-e interaction. They correspond to the diagonal components of the dynamical dielectric function matrix and the MO conductivity matrix. By considering the current-current correlation induced by e-e interaction in the Kubo formula \cite{Zhao20,Tabert13,Gusynin07,Zhou21}, we can calculate the transverse or Hall MO conductivity $\sigma_{xy}(\omega)$, which is the off-diagonal component in MO conductivity matrix. From $\sigma_{xy}(\omega)$, we can obtain the off-diagonal component of the dynamical dielectric function matrix $\epsilon_{xy}(\omega)$ by using the relationship: $\epsilon_{xy}(\omega)=\kappa+i\sigma_{xy}(\omega)/(\epsilon_0\omega)$ with $\epsilon_0$ being the vacuum dielectric constant.
However, the calculation of $\sigma_{xy}(\omega)$ by using the Kubo formula requires the considerable further analytical and numerical work and, therefore we do not attempt it in the present study. Furthermore, at present most experiments use circularly polarized light to characterize the ML-TMDs. The left-handed circularly polarized (LHCP) and right-handed circularly polarized (RHCP) MO conductivities $\sigma_\pm(\omega)$ can be obtained via the relation: $\sigma_{\pm}(\omega)=\sigma_{xx}(\omega) \pm i\sigma_{xy}(\omega)$.

\section{Results and discussions}
\label{sec:results}

\begin{figure*}
\includegraphics[width=13.5cm]{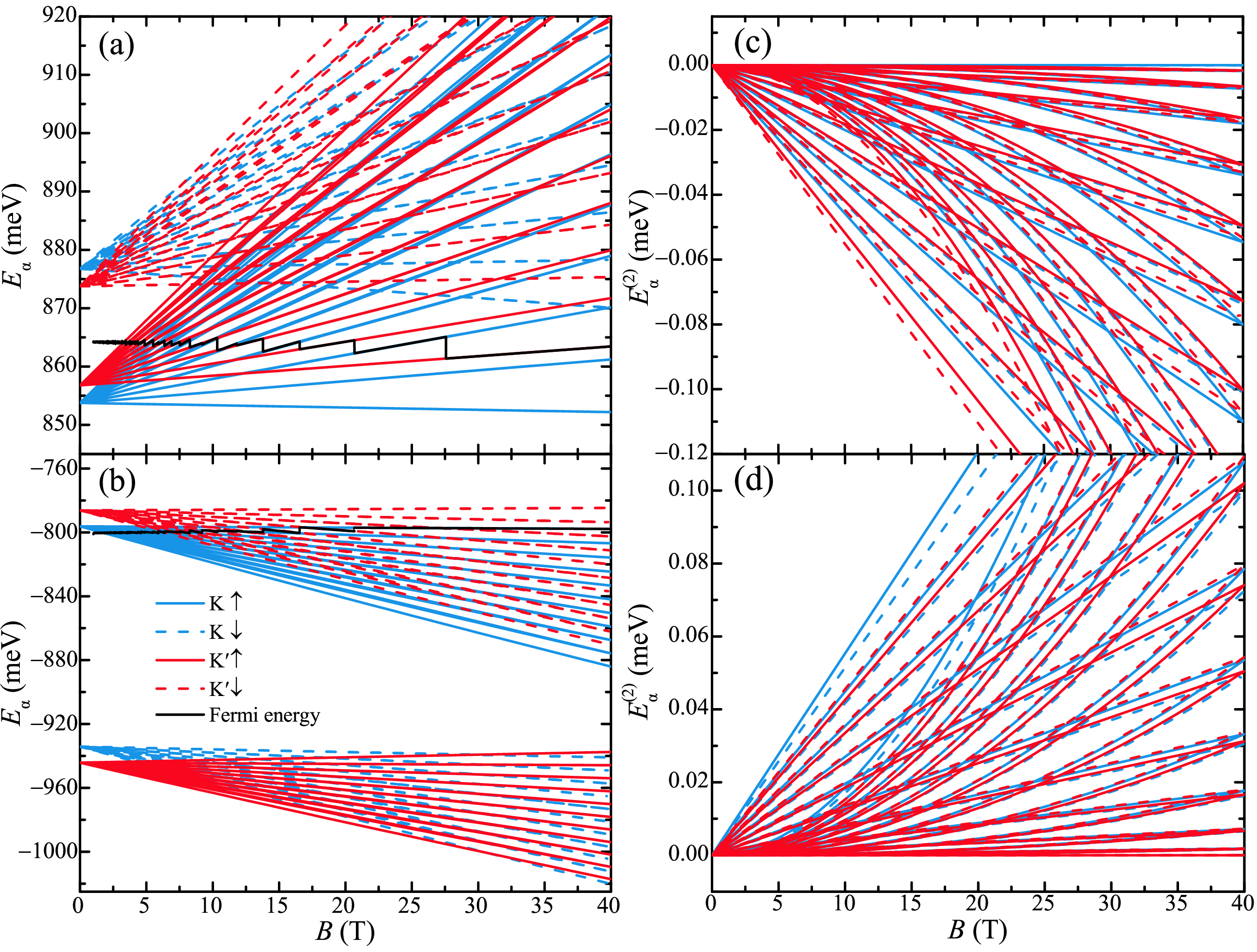}
\caption{The LLs $E_\alpha$ (left) and the second-order correction $E_\alpha^{(2)}$ (right) induced by trigonal warping, and Fermi energy (black curves) of $n$- and $p$-type ML-MoS$_2$
as a function of magnetic field strength $B$ for fixed carrier density
$n_\beta=2\times10^{12}$ cm$^{-2}$, electrical field strength $E_z$ = 35.25 mV/$d$, effective Zeeman
fields $B_c=10$ meV and $B_v=5$ meV. (a) and (c) as well as (b) and (d)
correspond to the conduction and valence bands, respectively. The spin up/down
LLs in $K$ and $K'$ valleys are indicated.}\label{fig1}
\end{figure*}

In this study, we consider $n$- and $p$-type ML-TMDs placed on
a ferromagnetic substrate where the proximity-induced exchange interaction
is presented. For the calculations, we consider the case of a low temperature
$T=4.2$ K, where the scattering mechanism such as the electron-phonon coupling
can be safely ignored and the inelastic electronic scattering for inter-level
electronic transition is mainly induced by electron-photon interaction. We take
ML-MoS$_2$ an example for the present study. The following material parameters
for ML-MoS$_2$ \cite{Xiao12,Liu13,Kormanyos14,Jena17} are taken in our numerical
calculations: $v_\mathrm{F}=5.3\times 10^5$ m/s, $\Delta=0.83$ eV, $\lambda_c=-1.5$ meV,
$\lambda_v =$ 74 meV, $2d=1.561 ${\AA}, g$_v=$3.57, and g$'_s = $0.21. Furthermore, we use the Lorentzian
type of LL broadening \cite{Xu96} to replace the $\delta$-function in Eq. \eqref{eq22} through
$\delta(x)\rightarrow (\Gamma/\pi)/(x^2+\Gamma^2)$ with $\Gamma$ being the width
of the broadened energy states of the LLs \cite{Tahir16,2Tahir16}. $\Gamma$ can
be approximately evaluated via \cite{Ando82,Wang12PRB}: $\Gamma=\alpha_0 B^{1/2}$
with a unit of meV when taking $B$ with a unit of Tesla. In the present study,
we take the relatively small values in order to see more clearly the basic features
of the MO absorption spectra in $n$- and $p$-type ML-MoS$_2$ in the presence
of proximity-induced exchange interaction and external electrical field.

In Fig. \ref{fig1}, we plot the LLs $E_\alpha$, second-order correction $E_\alpha^{(2)}$
relative to the LLs induced by trigonal warping, and Fermi energy for $n$- and $p$-type
ML-MoS$_2$ as a function of magnetic field strength $B$ for fixed carrier
densities $n_\beta$, electrical field strength $E_z$, and effective Zeeman fields
$B_c$ and $B_v$. We can see that the second-order corrections of LLs are
approximately $10^4$ times smaller than the LLs in the absence of trigonal
warping term. Thus the impact of trigonal warping on the LLs and the position
of MO absorption peaks is indeed very small. Meanwhile, the trigonal warping can
induces additional allowed transition channels $n\rightarrow$ $n$ $\pm$ 2
and $n\rightarrow$ $n$ $\pm$ 4 but the strengthes of new MO absorption peaks
would be relatively weak and we can safely ignore it in our following calculations \cite{Rose13}.
The energies of LLs in conduction (valence) band increase
(decrease) linearly with increasing magnetic field $B$, which is the common
feature of LLs in 2DEG system with parabolic band structure under the action
of the perpendicular magnetic field \cite{Tahir16,2Tahir16,Hien20}. From
Fig. \ref{fig1}, we can see that the LLs in conduction (valence) band in
valley $K^\prime$ are higher (lower) than those in valley $K$. This is
because in the presence of proximity-induced exchange interaction and when
$B\to 0$, the valley degeneracy in the conduction and valence bands can been
lifted \cite{Zhao20}. In $n$-type ML-MoS$_2$ with a fixed electron density
$n_+=2\times10^{12}$ cm$^{-2}$, the electrons mainly occupy the spin-up LLs,
where the full spin-polarization can be always achieved even in large magnetic
fields. In contrast, as can be seen from Fig. \ref{fig1}(b), in $p$-type ML-MoS$_2$
with a low hole density the holes mainly occupy the spin-down LLs, where the
full spin-polarization can be achieved in small magnetic fields. We also noted
that Ref. \cite{Tahir16} investigated the effect of spin and valley Zeeman
fields on magneto-transport properties of ML-MoS$_2$ and the lifting of valley
degeneracy occurs when the spin and valley Zeeman fields are presented. However,
the effective Zeeman fields originated from the proximity-induced exchange interaction
can largely enhance the breaking of the spin and valley degeneracies in the
presence of a magnetic field which can introduce the large spin and valley
polarizations.

\begin{figure}[t]
\includegraphics[width=8.2cm]{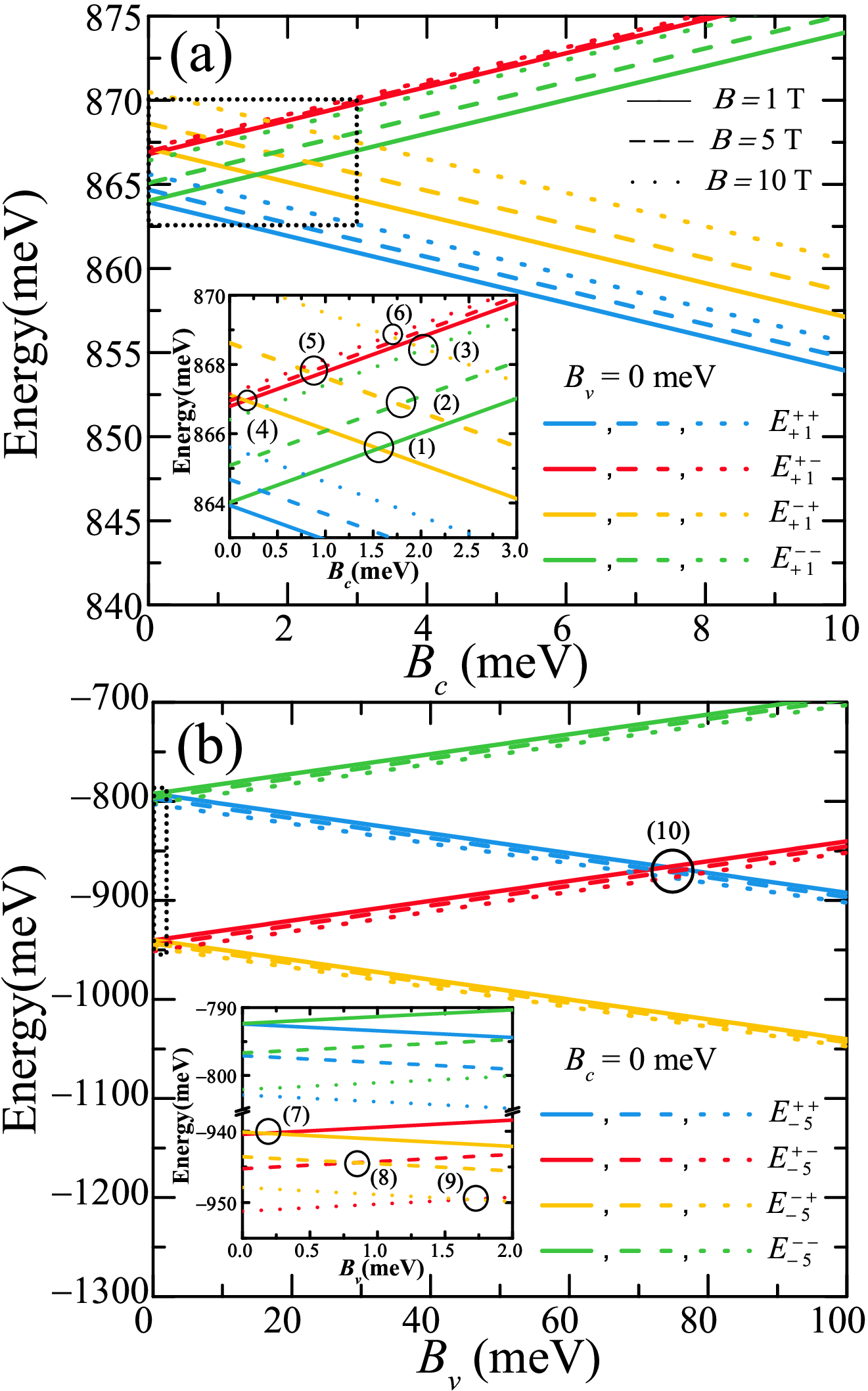}
\caption{The LLs $E_{\beta n}^{+ +}$ (blue), $E_{\beta n}^{+ -}$ (red),
$E_{\beta n}^{- +}$ (yellow), and $E_{\beta n}^{- -}$ (green) in
ML-MoS$_2$ in (a) conduction band with LL index $n=1$ as a function of
effective Zeeman fields $B_c$ ($B_v=0$ meV) and (b) valence band with
LL index $n=5$ as a function of effective Zeeman fields $B_v$ ($B_c=0$ meV)
at the fixed electrical field strength $E_z$ = 35.25 mV/$d$ for different
magnetic field strengths $B=1$ (solid line), $B=5$ (dashed line),
and $B=10$ T (dotted line), respectively. The insets in (a) and (b) are the
zooms in of the dotted rectangle areas.}\label{fig2}
\end{figure}

\begin{figure}[t]
\includegraphics[width=8.0cm]{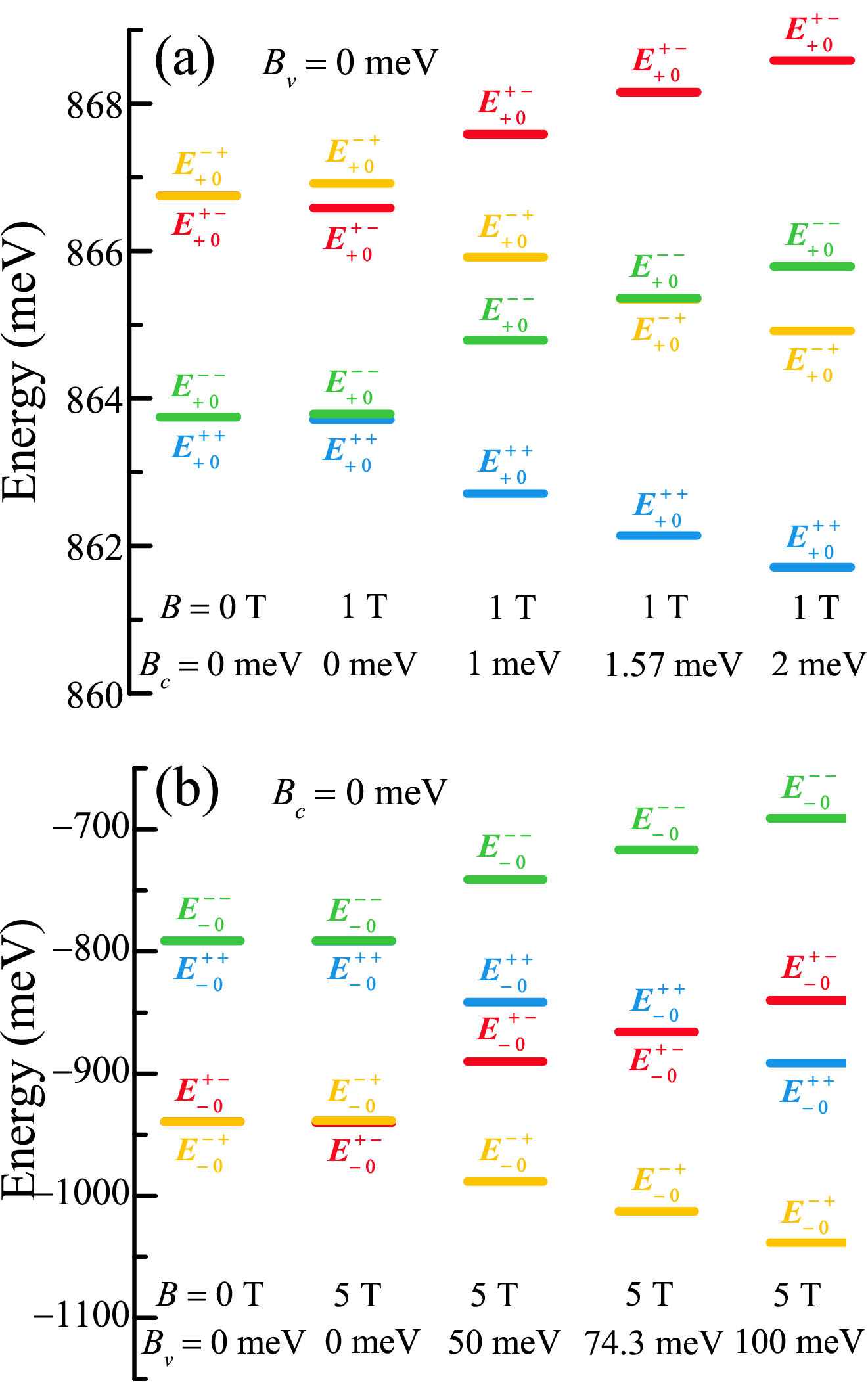}
\caption{Schematic illustration of the LLs structure in ML-MoS$_2$ with
LL index $n=0$ at the fixed external electrical field strength $E_z$ = 35.25 mV/$d$
for different magnetic fields $B$, effective Zeeman fields $B_c$ and
$B_v$. The upper panel (a) and lower panel (b) correspond to the conduction
and valence bands, respectively.}\label{fig3}
\end{figure}

\begin{figure}[t]
\includegraphics[width=8.2cm]{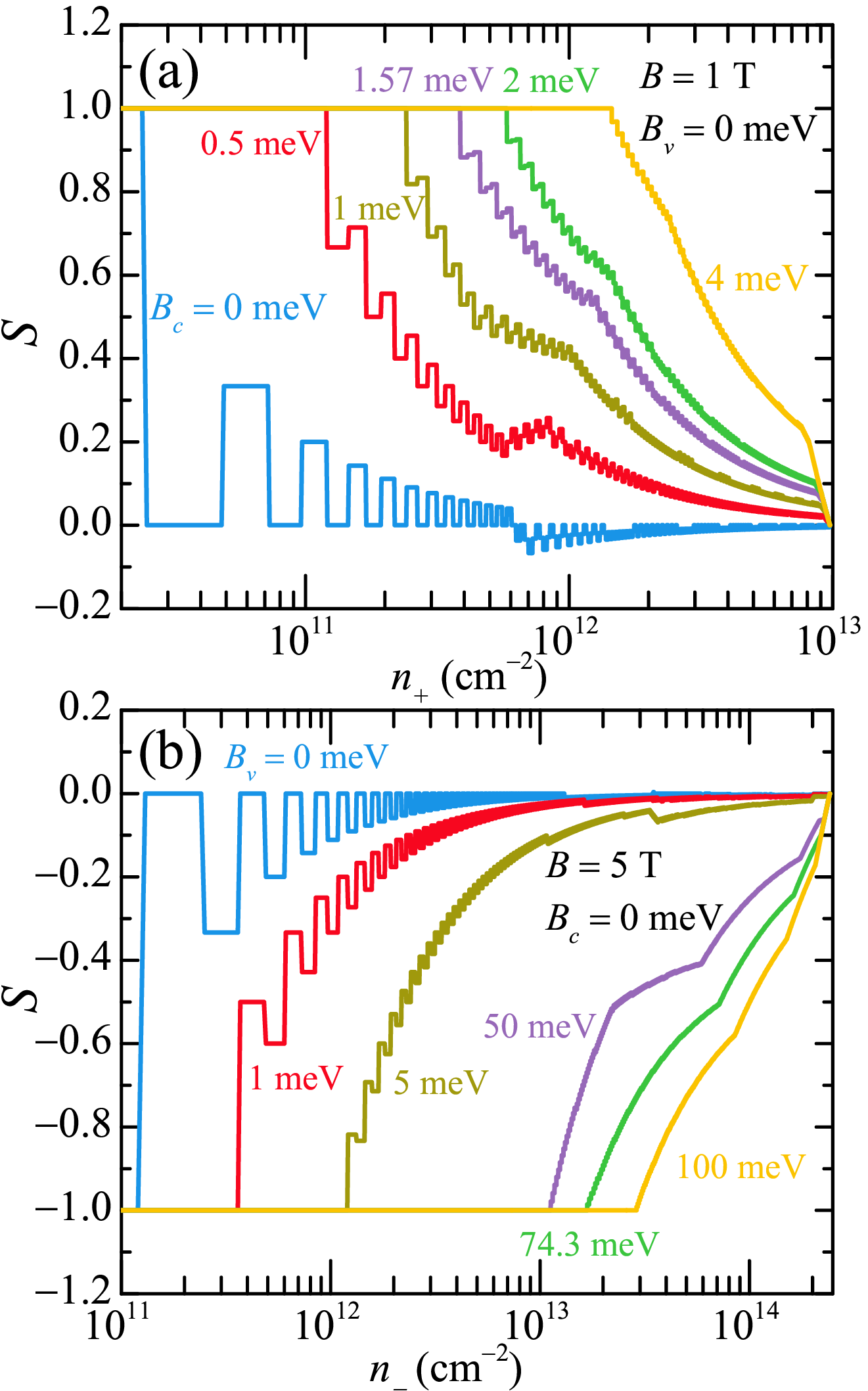}
\caption{The spin polarization $\mathcal{S}$ of $n$-type and $p$-type ML-MoS$_2$
as a function of electron/hole density $n_{\beta}$ at the fixed temperature
$T = 0$ K, external electrical field strength $E_z$ = 35.25 mV/$d$,
and magnetic field $B$ for different effective Zeeman fields $B_c$ and $B_v$. (a) and (b)
correspond to the conduction and valence bands, respectively.}\label{fig4}
\end{figure}

In Fig. \ref{fig2}, we plot the LLs of ML-MoS$_2$ in conduction with
LL index $n$ = 1 and valence bands with LL index $n$ = 5 as a function
of experienced effective Zeeman field $B_c$ ($B_v$ = 0 meV) in (a) and
$B_v$ ($B_c$ = 0 meV) in (b) at fixed electrical field strength $E_z$ for
different magnetic field strengths $B$ to clearly show the influence
of proximity-induced exchange interaction on the LLs spectrum of
ML-MoS$_2$. We can see that the energies of LLs increases with $B_c$ ($B_v$)
for spin down subbands and decreases with $B_c$ ($B_v$) for spin up subbands.
For a specific experienced effective Zeeman field, we notice that
there exist LLs degeneracy as shown by the black circles indicated by
numbers $(1)\sim(10)$. At the intersection shown by circle $(1)$, the
spin up electrons and the spin down electrons in $K^\prime$ valley are
degenerate when $B_c=$ 1.54 meV and the energies of spin subbands are
reversed with $B_c>$ 1.54 meV. Similarly, the same features occur at
circles $(2)\sim(10)$ which indicate that the spin states can be
effectively tuned by the proximity-induced interaction. Compared to the
conduction band, the valence band requires a larger experienced effective
Zeeman fields to achieve the reversals. In addition, with increasing the
 magnetic field strength $B$, the locations of the intersections are
also shifted. In Fig. \ref{fig3}, we show the schematic illustration of
the LLs structure of ML-MoS$_2$ with LL index $n=0$ at fixed external
electrical field strength $E_z$ for different magnetic fields $B$, experienced effective
Zeeman fields $B_c$ and $B_v$. From the first two columns in Fig. \ref{fig3}(a),
we can see that magnetic field can break the energy degeneracy in two
valleys. However, this behavior is not obvious in Fig. \ref{fig3}(b)
because it needs larger magnetic field in valence band. We find that the
LLs splitting increases with increasing the experienced effective Zeeman
field $B_c$ ($B_v)$. In Fig. \ref{fig3}(a), with $B_c = 1$ meV, the LLs
energy $E_{+ 0}^{+ +} < E_{+ 0}^{- -} < E_{+ 0}^{- +} < E_{+ 0}^{+ -}$,
with $B_c = 1.57$ meV, the LLs $E_{+ 0}^{- -}$ and $E_{+ 0}^{- +}$ are
degenerate, with $B_c = 2$ meV, the LLs energy
$E_{+ 0}^{+ +} < E_{+ 0}^{- +} < E_{+ 0}^{- -} < E_{+ 0}^{+ -}$,
we can see that spin indexes of the spin subbands of LLs are
interchanged at $B_c>$ 1.54 meV. Similarly, the same features
occur at valence bands in Fig. \ref{fig3}(b). We can obtain that
proximity-induced interaction can reverse the spin indexes of the
spin subbands of LLs. Moreover, at a magnetic filed $B=1$ T, the LL
$E_{+0}^{+-}$ are lifted with the energy higher than LL $E_{+0}^{-+}$ after
applying an experienced effective Zeeman field $B_c=1$ meV. Thus,
the proximity-induced interaction and magnetic field can effectively tune
the energies and degeneracy of LLs in ML-MoS$_2$.

In Fig. \ref{fig4}, we plot the spin polarization $\mathcal{S}$ of $n$-type and
$p$-type ML-MoS$_2$ as a function of total electron (hole) density $n_{\beta}$
at fixed external electric field strength, and magnetic field for different experienced effective
Zeeman fields. In general, we can see that the absolute value of spin polarization
$|\mathcal{S}|$ decreases with increasing carrier density. We can explain the
trend of spin polarization in Fig. \ref{fig4} with the help of Fig. \ref{fig3}.
For example, with $B_c = 1$ meV in Fig. \ref{fig3}(a), the LLs energies have
the feature of $E_{+ 0}^{+ +} < E_{+ 0}^{- -} < E_{+ 0}^{- +} < E_{+ 0}^{+ -}$ .
With increasing carrier density, electrons would first fill into the lower spin
up subbands $E_{+ 0}^{+ +} $ with the 100\% spin polarization, and then fill into
spin down subbands $E_{+ 0}^{- -}$ with decreasing the spin polarization. We can
see that spin polarization has a tendency to increase because the spin up subband
$E_{+ 0}^{- +}$ are filled at this time. With the occupation of electrons to
spin down subband $E_{+ 0}^{+ -}$, the spin polarization will gradually decreases to zero.
For low carrier density, the spin polarization is tortuous because of the
overlap between Landau levels. The increasing of the experienced effective
Zeeman fields $B_c$ in $n$-type and $B_v$ in $p$-type MoS$_2$ can sustain a full
spin polarization for a large range of carrier density. For a fixed carrier
density, we find that the spin polarization would increases with increasing the
effective Zeeman fields. In addition, we find that a small effective Zeeman field
factors $B_c$ can achieve a large spin polarization compared to previous studies
where a large Rashba parameter is required to achieve spin polarization \cite{Zhao20}.
From Fig. \ref{fig4}, we obtained that the proximity-induced exchange interaction
can effectively tuned the spin polarization of ML-MoS$_2$ under a magnetic
field. One can acquire a large spin polarization by the effective Zeeman fields
from the proximity-induced exchange interaction.

\begin{figure}[t]
\includegraphics[width=8.2cm]{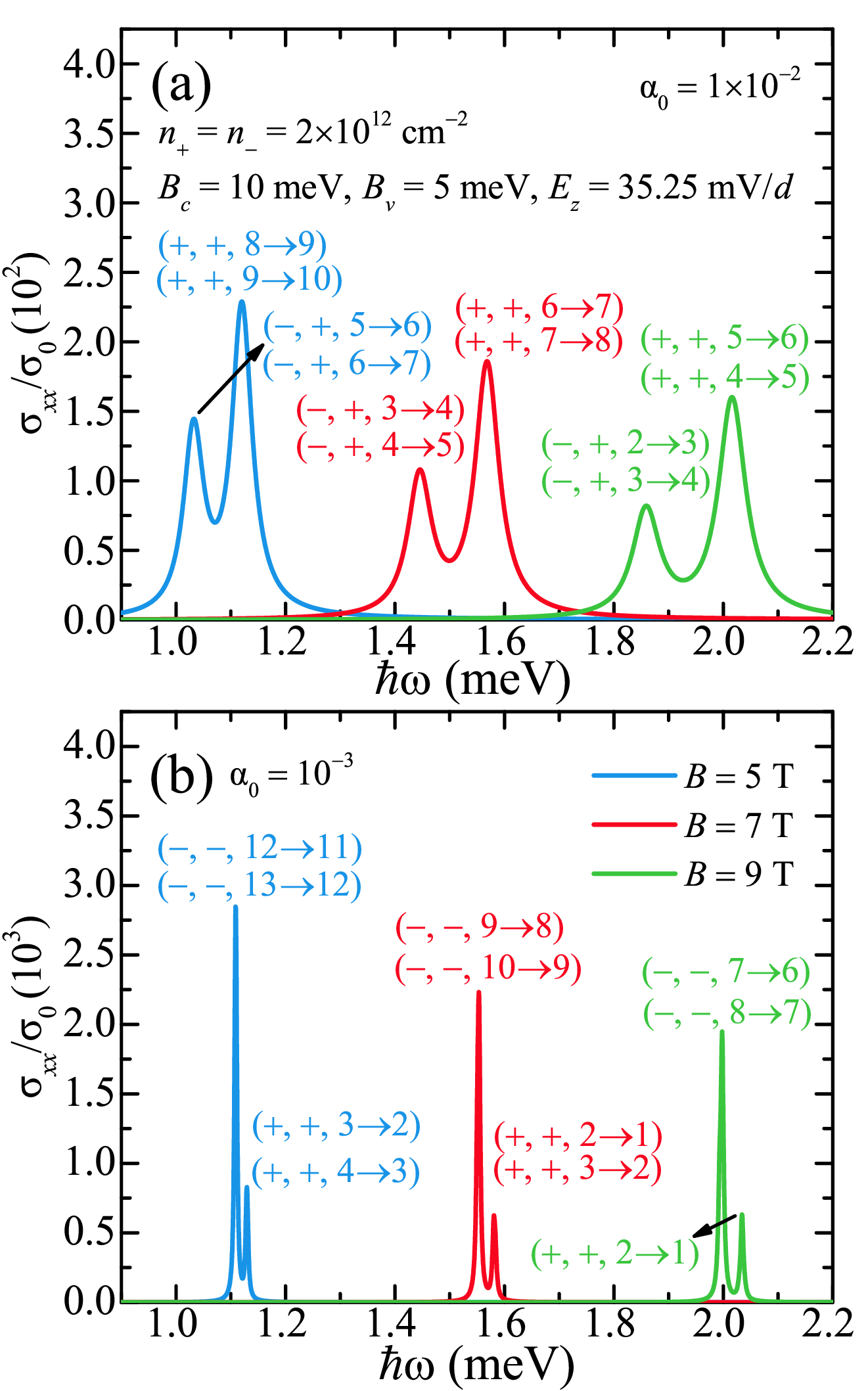}
\caption{Magneto-optical conductivity induced via the electronic transition
within the conduction bands in (a) and valence bands in (b) in ML-MoS$_2$ as
a function of radiation photon energy $\hbar\omega$ at the fixed carrier
densities $n_\beta = 2\times 10^{12}$ cm$^{-2}$, external electrical field strength $E_z$ = 35.25 mV/$d$,
effective Zeeman field factors $B_c=10$ meV and $B_v=5$ meV for different magnetic
fields $B=5$ (blue curve), $B=7$ (red curve), and $B=9$ T (green curve),
respectively. The electronic transition channels corresponding to the absorption
peaks are indicated by $(\zeta,s,n\rightarrow n')$. Here $\sigma_0=e^2/h$.}\label{fig5}
\end{figure}

\begin{figure}[t]
\includegraphics[width=8.2cm]{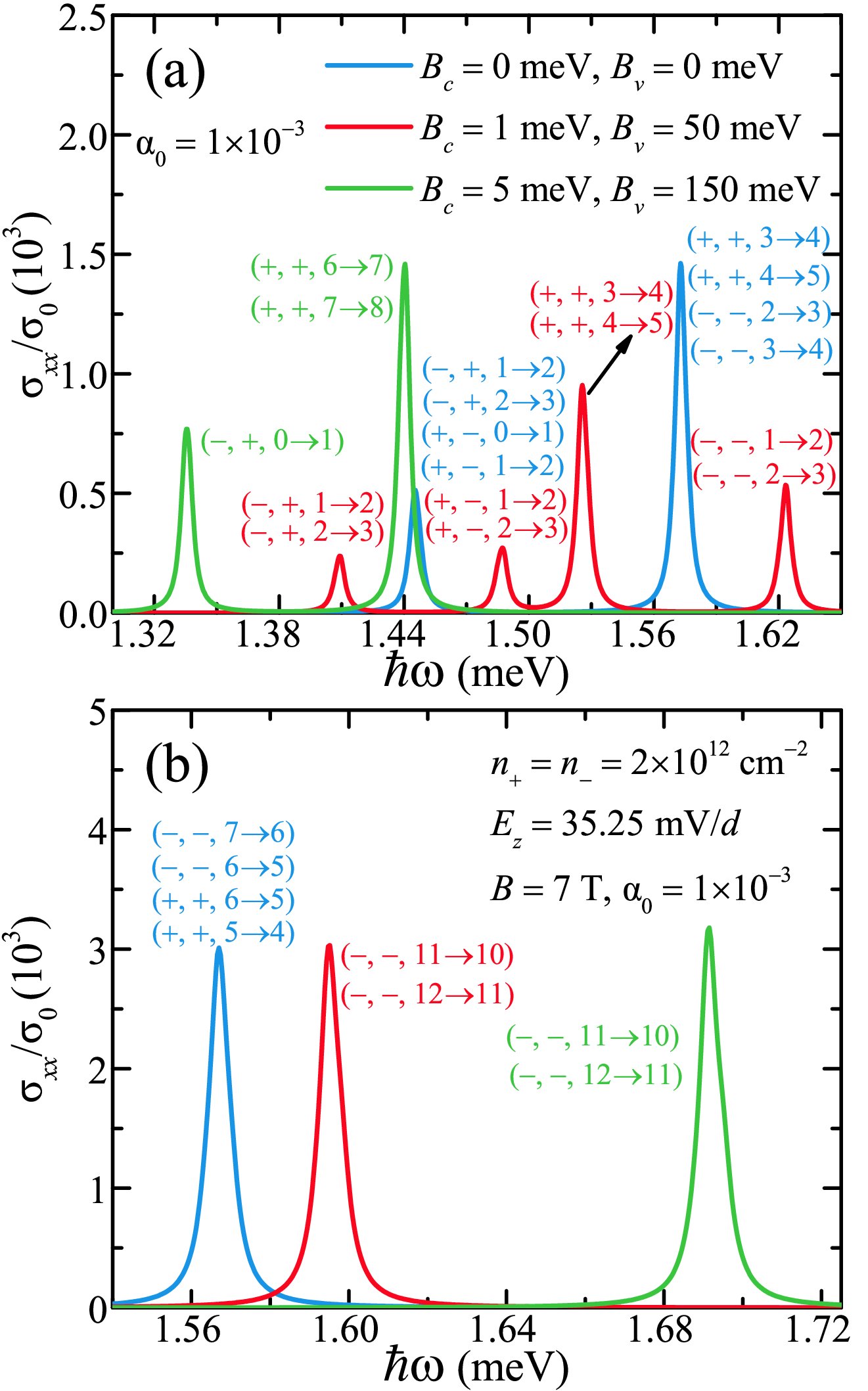}
\caption{Magneto-optical conductivity induced by intra-band electronic transition
within (a) the conduction band and (b) the valence band as a function of radiation photon
energy $\hbar\omega$ at the fixed carrier density $n_\beta$, external electrical field strength $E_z$ = 35.25 mV/$d$
and magnetic field $B=7$ T for different effective Zeeman fields $B_c=0$ meV, $B_v=0$ meV
(blue curve), $B_c=1$ meV, $B_v=50$ meV (red curve), and $B_c=5$ meV, $B_v=150$
meV (green curve), respectively.}\label{fig6}
\end{figure}

\begin{figure}[t]
\includegraphics[width=8.2cm]{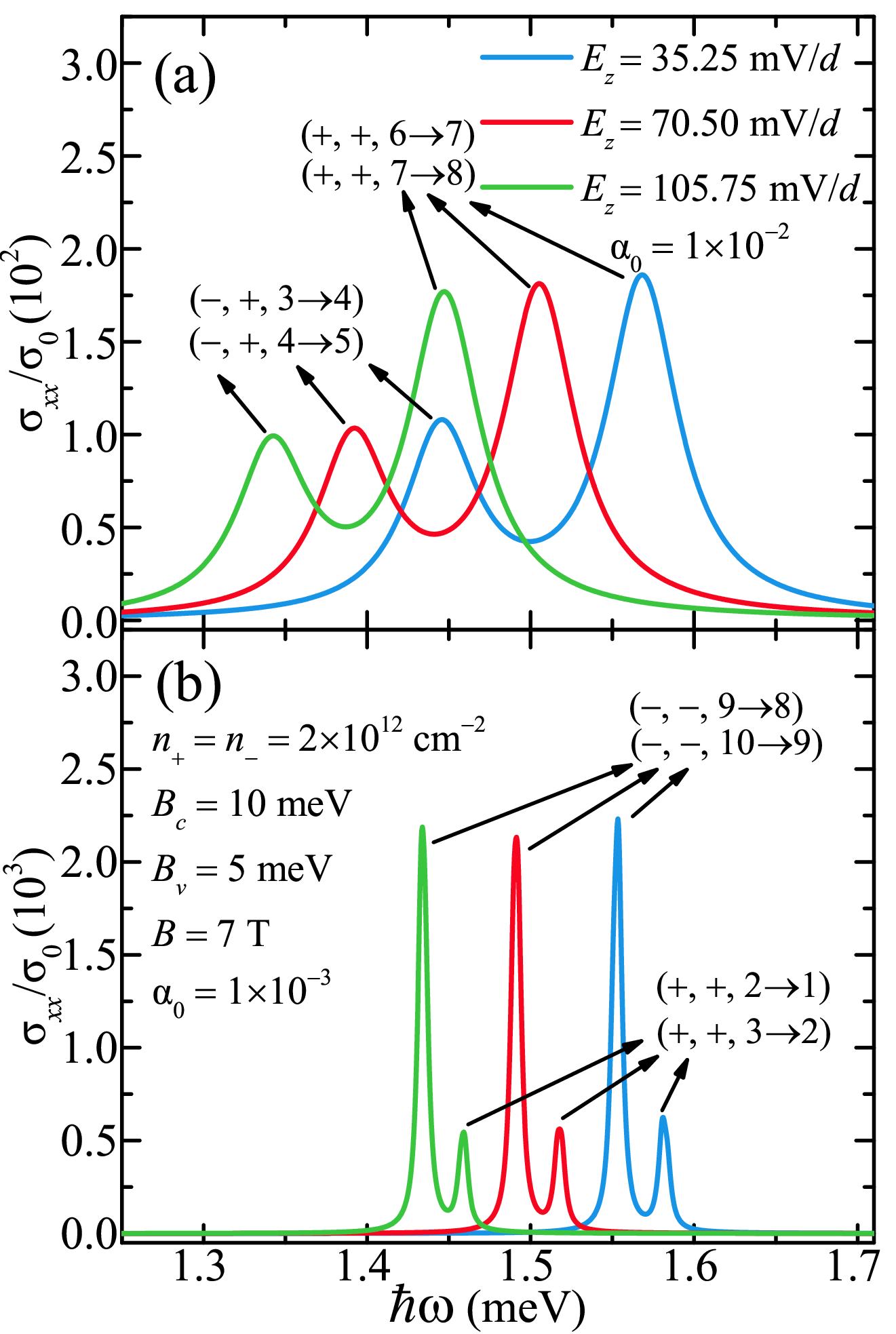}
\caption{Magneto-optical conductivity induced by intraband electronic transitions
within (a) the conduction band in $n$-type ML-MoS$_2$ and (b) the valence
band in $p$-type ML-MoS$_2$ as a function of radiation photon energy $\hbar\omega$
at the fixed carrier density $n_\beta=2.0\times10^{12}$ cm$^{-2}$,
effective Zeeman fields $B_c=10$ meV and $B_v=5$ meV, and magnetic field $B=7$ T
for different external electrical field strengths $E_z$ = 35.25 mV/$d$ (blue curve),
$E_z=70.5$ mV/$d$ (red curve), and $E_z=105.75$ mV/$d$ (green curve),
respectively.}\label{fig7}
\end{figure}

\begin{figure}[t]
\includegraphics[width=8.2cm]{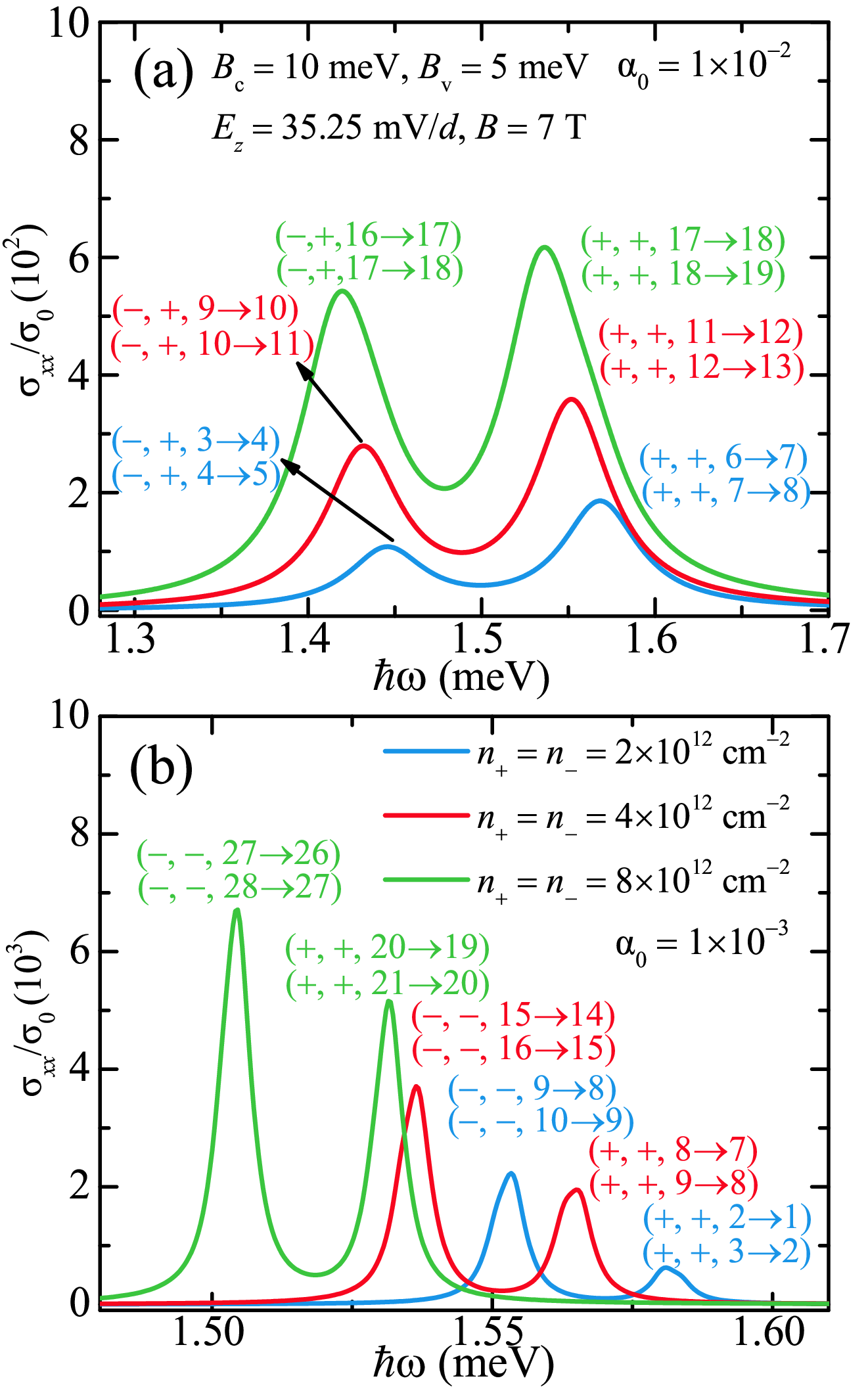}
\caption{Magneto-optical conductivity induced by intra-band electronic transitions
within (a) the conduction band in $n$-type ML-MoS$_2$ and (b) the valence
band in $p$-type ML-MoS$_2$ as a function of radiation photon energy $\hbar\omega$
at the fixed effective Zeeman fields $B_c=10$ meV and $B_v=5$ meV,
external magnetic field $B=7$ T and external electrical field strength $E_z$ = 35.25 mV/$d$
 for different carrier densities $n_\beta=2.0\times10^{12}$ cm$^{-2}$ (blue curve),
$n_\beta=4.0\times10^{12}$ cm$^{-2}$ (red curve), and
$n_\beta=8.0\times10^{12}$ cm$^{-2}$ (green curve), respectively.}\label{fig8}
\end{figure}
The MO conductivity $\sigma_{xx}(\omega)$ induced via the electronic transitions
within the conduction and valence bands as a function of radiation photon
energy $\hbar\omega$ at the fixed carrier densities, external electric field strength,
and effective Zeeman field factors for different magnetic field strengths are
shown in Fig. \ref{fig5}. As can be seen, there are two absorption peaks in MO
conductivity spectrum in the THz regime for both $n$- and $p$-type ML-MoS$_2$
with different magnetic fields corresponding to
transition energies $\hbar \omega_{1} = |E_{\beta n}^{\zeta s}
-E_{\beta n\pm1}^{\zeta s}| \approx \hbar^2\omega^2_c/(2\Delta_{\zeta s})$.
This behavior differs from previous report where only one absorption peak was
observed \cite{Huong20,Nguyen18}. Such an effect is a consequence of the breaking
of the band symmetry in $K$ and $K^\prime$ valleys due to the presence of the
proximity-induced exchange interaction. We note that two electronic transition
channels $(\zeta,s,n\rightarrow n')$ contribute to each absorption peak because
$\hbar \omega_{1}$ depends less on LLs index $n$ and the allowed electronic
transitions occur only near the Fermi Level (or chemical potential at a finite
temperature). For $n$-type ML-MoS$_2$ in Fig. \ref{fig5}(a), the two absorption
peaks are due to the electronic transitions for spin up electrons in valley
$K^\prime$ and valley $K$, respectively. In Fig. \ref{fig5}(b), the first absorption
peak in $p$-type ML-MoS$_2$ is contributed by the electronic transitions for spin
down holes in valley $K^\prime$ and the second one relates to the electronic
transitions for spin up holes in valley $K$. These spin and valley dependent
electronic transition channels are attributed to the spin and valley polarization
caused by the experienced effective Zeeman fields. The energy spacing between two
absorption peaks increases with the magnetic field strength, indicating that the
MO absorption peaks for different transition channels between LLs can be more
obviously observed in stronger magnetic field. With increasing magnetic field
strength, the absorption peaks in MO conductivity show a blue-shift due to the
increase in energy spacing between LLs as shown in Fig. \ref{fig1}.

In Fig. \ref{fig6}, we plot the MO conductivity induced by electronic
transitions within the conduction and valence bands of $n$-type and
$p$-type ML-MoS$_2$ as a function of photon energy $\hbar\omega$ at the fixed
carrier densities, external electric field and magnetic fields for different
experienced effective Zeeman fields. With increasing experienced
 effective Zeeman fields $B_c$ and $B_v$, the MO absorption peaks have
blue-shifts for spin down subbands and red-shifts for spin up subbands
both in $n$-type and $p$-type ML-MoS$_2$. We note that
the effect of effective Zeeman fields $B_c$ and $B_v$ on LLs are
dependent upon the spin index $s$. Thus the electronic transition energies
$\hbar\omega_1 \approx \hbar^2\omega^2_c/(2\Delta_{\zeta s})$
increase for spin down subbands and decrease for spin up subbands
with decreasing the value of $B_c-B_v$. And when $B_c \neq 0$, there are
more MO absorption peaks, because the proximity-induced interaction can
break the band symmetry in $K$ and $K^\prime$ valleys. It should be noted
that the experienced effective Zeeman fields $B_c$ and $B_v$ are induced by the
proximity-induced interaction which can be different with the choosing of the
substrates \cite{Zhao20}. Therefore the MO conductivity
via the electronic transitions within the conduction and valence bands
in THz regime can be effectively tuned by the proximity-induced interaction.

\begin{figure}[t]
\includegraphics[width=8.6cm]{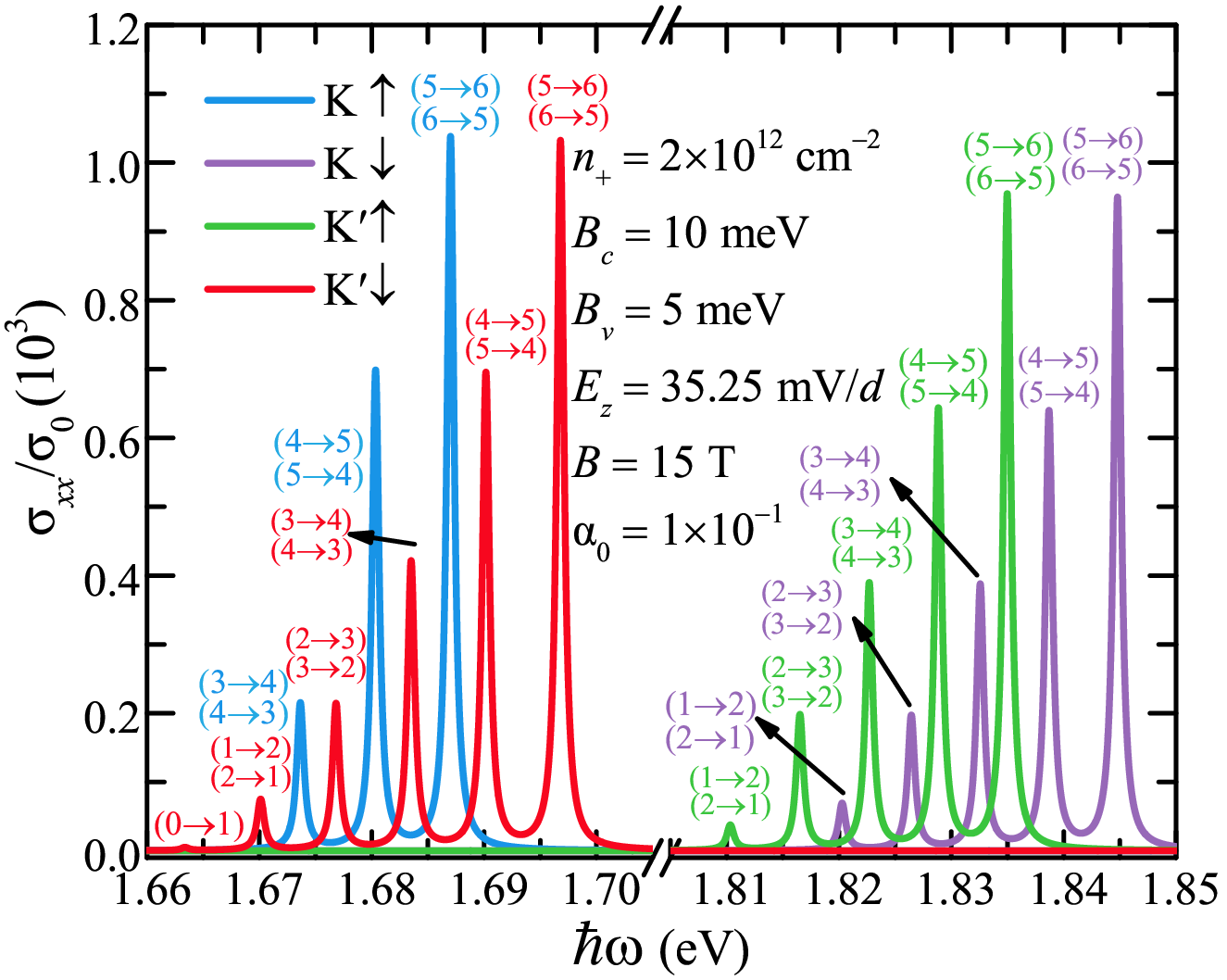}
\caption{Magneto-optical conductivity induced by inter-band electronic transitions
from valence band to conduction band in $n$-type ML-MoS$_2$ as a function of radiation
photon energy $\hbar\omega$ at the fixed electron density $n_+$ = 2 $\times 10^{12}$
cm$^{-2}$, external electrical field strength $E_z$ = 35.25 mV/$d$, effective Zeeman fields $B_c=10$ meV and $B_v=5$ meV, and magnetic field $B=15$ T. The electronic transition channels corresponding
to the absorption peaks are indicated by $(\zeta,s,n\rightarrow n')$.}\label{fig9}
\end{figure}

\begin{figure}[t]
\includegraphics[width=8.6cm]{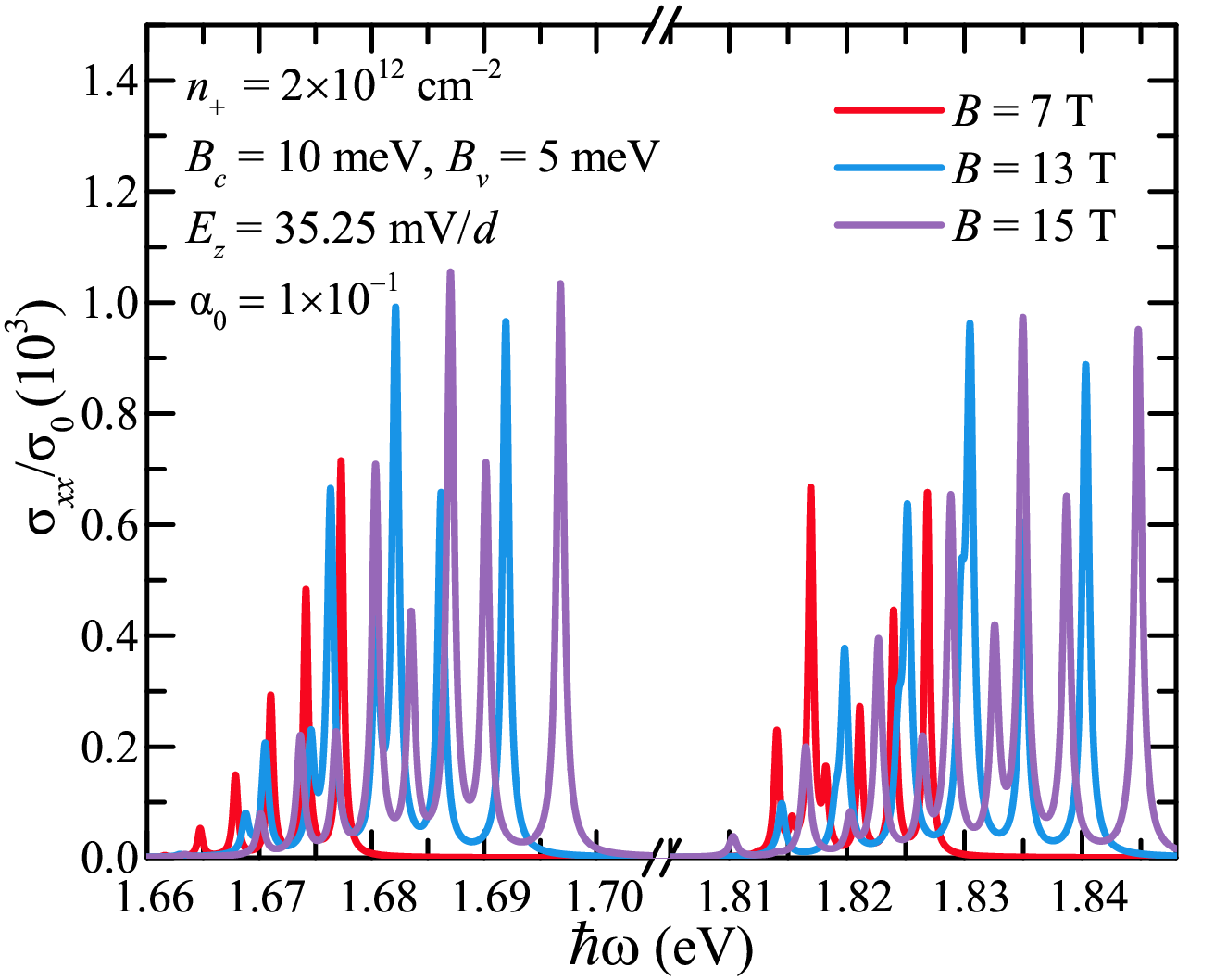}
\caption{Magneto-optical conductivity in $n$-type ML-MoS$_2$ as a function of
radiation photon energy $\hbar\omega$ with including the first six interband
electronic transition channels at the fixed electron density $n_+$ =2 $\times
10^{12}$ cm$^{-2}$, external electrical field strength $E_z$ = 35.25 mV/$d$, and effective Zeeman fields
$B_c=10$ meV and $B_v=5$ meV for different magnetic fields $B=7$ (red curve),
$B=13$ (blue curve), and $B=15$ T (purple curve), respectively.}\label{fig10}
\end{figure}

In Fig. \ref{fig7}, we show the MO conductivity induced by electronic
transitions within the conduction and valence bands for $n$-type and
$p$-type ML-MoS$_2$ as a function of radiation photon energy $\hbar\omega$
at the fixed carrier densities, magnetic field, experienced effective Zeeman
fields for different external electric field strengths. The increase in the external electric field
can lead to lower electronic transition energies
$\hbar\omega_1 \approx \hbar^2\omega^2_c/(2\Delta_{\zeta s})$
for both $n$-type and $p$-type ML-MoS$_2$. Therefore two MO absorption peaks
in Figs. \ref{fig7}(a) and \ref{fig7}(b) show red-shifts with increasing
$E_z$. Consequently, the tuning of the MO absorption can be effectively achieved
through applying a perpendicular electrical field. We also plot the MO
conductivity caused by electronic transitions within the conduction and
valence bands for $n$-type and $p$-type ML-MoS$_2$ as a function of radiation
photon energy $\hbar\omega$ at the fixed magnetic field, experienced effective
Zeeman fields and external electric field for different carrier densities in
Fig. \ref{fig8}. We can see that two absorption peaks also have red-shifts and
the hight of the absorption peaks increase with increasing the carrier density.
In Fig. \ref{fig8}, the allowed electronic transitions occur in high index LLs
at larger carrier density because the LLs with low index are fully occupied at
low temperature. As we known that the carrier density and chemical potential of
a 2DEG system can be tuned by, e.g., applying a gate voltage. Thus the MO
absorption in ML-MoS$_2$ in THz regime can also be effectively tuned with
varying the carrier density.

\begin{figure}[t]
\includegraphics[width=8.6cm]{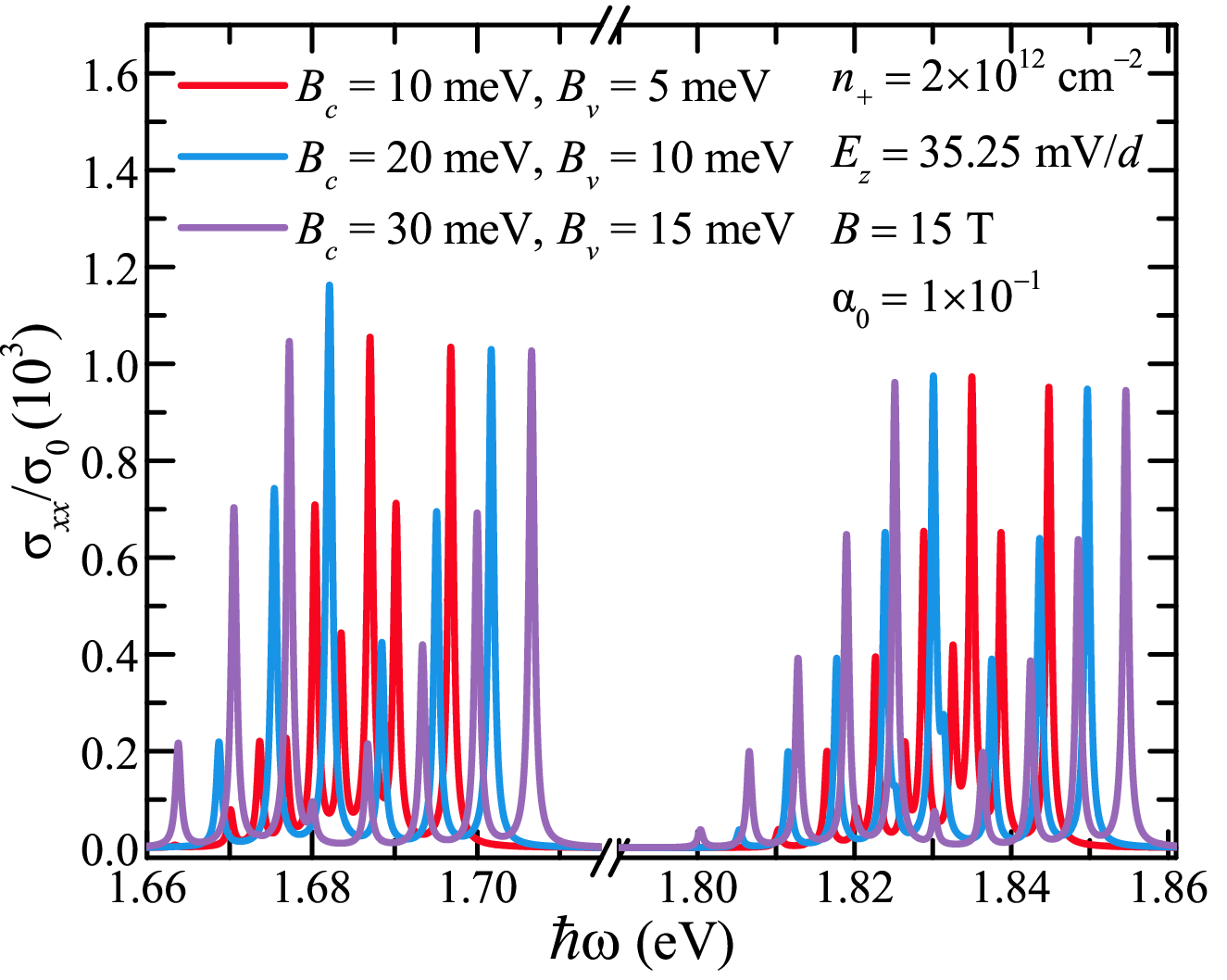}
\caption{Magneto-optical conductivity in $n$-type ML-MoS$_2$ as a function of
radiation photon energy $\hbar\omega$ with including the first six inter-band
electronic transition channels at the fixed electron density $n_+$ =2 $\times
10^{12}$ cm$^{-2}$, external electrical field strength
$E_z$ = 35.25 mV/$d$ and magnetic fields $B=15$ T
for different effective Zeeman fields $B_c=10$ meV, $B_v=5$ meV (red curve),
$B_c=20$ meV, $B_v=10$ meV (blue curve), and $B_c=30$ meV, $B_v=15$ meV (purple curve),
respectively.} \label{fig11}
\end{figure}

Different from electronic transitions within the conduction and/or valence
bands, which can have two MO absorption peaks in THz frequency regime, the
interband electronic transitions among the LLs in the conduction and valence
in $n$-type ML-MoS$_2$ can show a series of MO absorption peaks in visible
frequency range owing to the presence of the band gap. In Fig. \ref{fig9},
we plot the MO conductivity for each spin and valley contributed from the
first six interband electronic transition channels as a function of radiation
photon energy $\hbar\omega$ at the fixed effective Zeeman fields $B_c$ and $B_v$,
magnetic field $B$, external electrical field strength $E_z$, and electron density for an $n$-type
ML-MoS$_2$. The photon energies for inter-band transitions between conduction
and valence LLs are given by $\hbar\omega_2 = |E_{\beta n}^{\zeta s}
-E_{\beta^\prime n\pm1}^{\zeta s}|\approx(2n+1)\hbar^2\omega^2_c/(2\Delta_{\zeta s})
+2\Delta_{\zeta s}$, where $n=0, 1, 2, 3, \cdot\cdot\cdot$, resulting in a series of
absorption peaks in visible range. Besides, with increasing LL index
$n$, the position and the intensity of the absorption peak increases. These results
agree with the previous reports for MO response of ML-TMDs \cite{Hien20,Huong20}.

In Fig. \ref{fig10}, we show the total MO conductivity attributed
from the first six interband electronic transition channels as a function of
radiation photon energies $\hbar\omega$ at the fixed effective Zeeman fields
$B_c$ and $B_v$, external electric field \textcolor{blue}{ $E_z$}, and electron density with
different magnetic fields $B$ for an $n$-type ML-MoS$_2$. We note that the
energy for interband electronic transition $\hbar\omega_2$ increases with the
magnetic field strength. Thus the MO absorption peaks move to higher
frequency regime with increasing magnetic field strength. This effect
can be observed experimentally via, e.g., the visible MO-spectroscopy
measurement \cite{Delhomme19}.

The total MO conductivity attributed from the first six interband
electronic transition channels as a function of radiation photon energies
$\hbar\omega$ at the fixed external electric field \textcolor{blue}{$E_z$}, electron density $n_+$,
and magnetic field $B$ for different effective Zeeman fields $B_c$ and
$B_v$ for an $n$-type ML-MoS$_2$ is shown in Fig. \ref{fig11}. With increasing the
value of $B_c-B_v$, we can observe that the MO absorption peaks have both of the
blue-shifts and the red-shifts. This phenomena is due to the contributions
from different spin subbands at $K$ and $K^\prime$ valleys. We can see that
the electronic transition energy $\hbar\omega_2$ increases with $B_c-B_v$
for $s=-1$ and decreases with $B_c-B_v$ for $s=+1$. Thus the blue-shifts are
attributed to the electronic transitions of spin-down electrons and the
red-shifts are attributed to electronic transitions of spin-up electrons.

\begin{figure}[t]
\includegraphics[width=8.6cm]{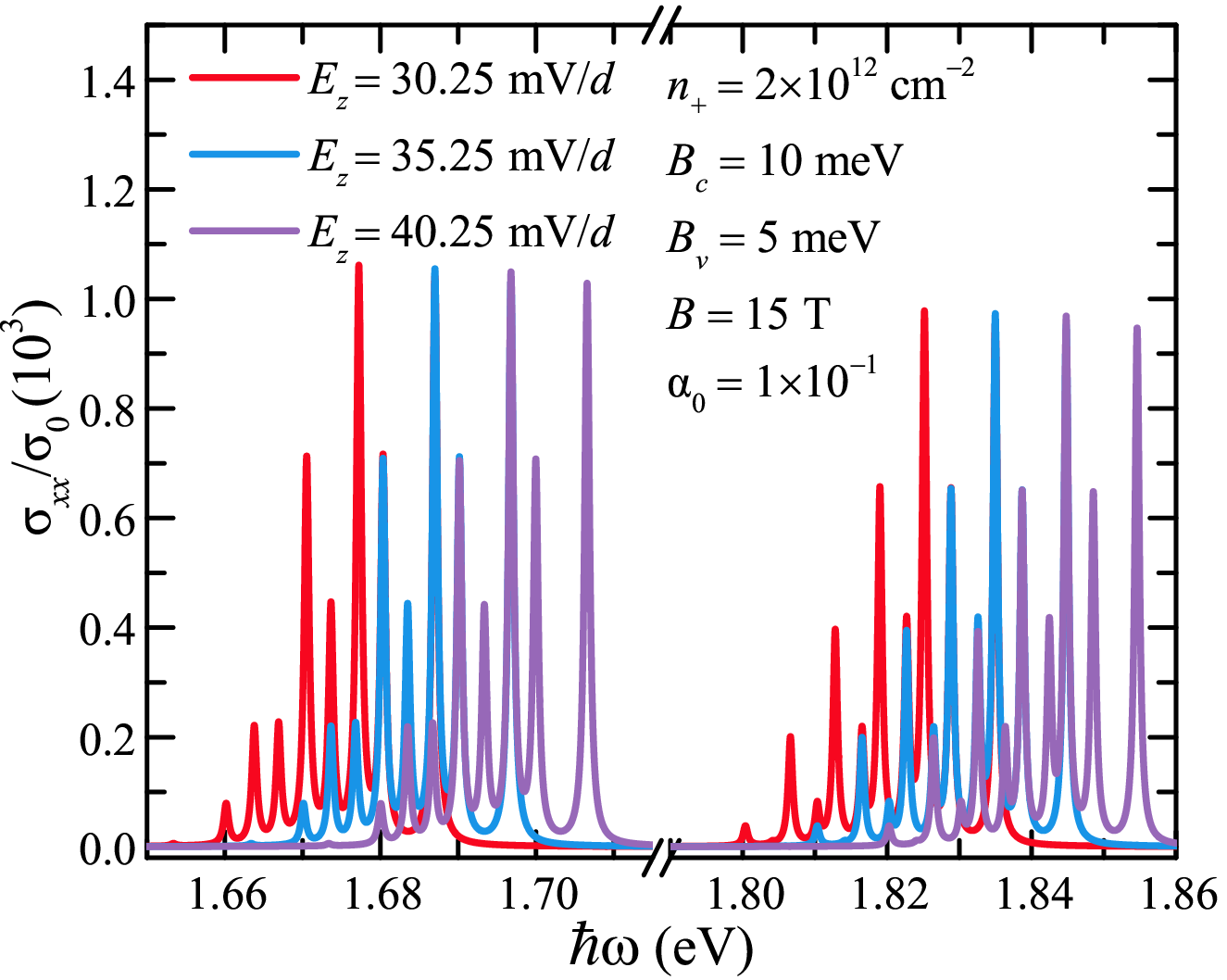}
\caption{Magneto-optical conductivity as a function of radiation photon energy
$\hbar\omega$ with including the first six interband electronic transition
channels at a fixed electron density $n_+$ =2 $\times 10^{12}$ cm$^{-2}$,
effective experienced Zeeman fields $B_c=10$ meV and $B_v=5$, magnetic field
$B=15$ T for different external electrical field strengths $E_z$ = 30.25 mV/$d$ (red curve), $E_z=35.25$
mV/$d$ (blue curve), and $E_z=40.25$ mV/$d$ (purple curve), respectively.}\label{fig12}
\end{figure}

In Fig. \ref{fig12}, we show the total MO conductivity induced by the first
six interband electronic transitions channels as a function of radiation
photon energy $\hbar\omega$ at the fixed electron density, magnetic field
strength and effective Zeeman fields $B_c$ and $B_v$ for different external
electric field strengths $E_z$. We find that the MO absorption peaks have blue-shifts with increasing external electric field strength $E_z$. This is understandable with the help of
Eq. \eqref{egvalue}. The LLs in conduction/valence are higher/lower with
increasing external electric field. Hence, the photon energy required for electronic
transitions from valence LLs to conduction LLs increases with $E_z$. By the
way, the external electrical field can be easily applied and tuned through a
gate voltage in a sample device setup. Thus we can predict that the electrically
tunable MO absorptions can be achieved experimentally.

In this study, we theoretically calculate the longitudinal MO conductivity
for ML-MoS$_2$ in the presence of proximity-induced exchange interaction
and external electrical field with the framework of RPA dielectric function.
The intraband and interband electronic transitions can be accompanied by
the absorption of THz and visible photons. The proximity-induced exchange
interaction can breaking the valley degeneracy and result in two MO absorption
peaks in THz regime for the intraband electronic transitions within the
conduction and/or valence LLs. The electronic transitions from valence LLs to
conduction LLs in $n$-type ML-MoS$_2$ results in a series of MO absorption peaks.
Moreover, we find that the proximity-induced exchange interaction, electrical
field, carrier density can significantly affect the MO absorption or conductivity
in ML-MoS$_2$.

As we know, in the presence of quantizing magnetic fields,
the quantum oscillations such as the Shubnikov-de Haas effect and the de
Haas-van Alphen effect can be observed in electron gas systems by
magneto-transport experiments at low-temperatures. In such case, the
oscillations of the longitudinal magneto-resistance/conductance can be
measured by varying the $B$ field, where the peak positions correspond
to the across of the Fermi energy over the LLs due to the opening up or
closing down of the electronic scattering channels. Therefore, the Fermi
energy in an electron gas can be evaluated via quantum oscillations observed
in magneto-transport measurements \cite{Soule64}. The peaks observed in MO
experiments for an electron gas in the presence of quantizing magnetic field
at a low-temperature, such as the results shown in this study, are the direct
consequence of electronic transitions from occupied LLs to unoccupied or empty
LLs accompanied by the absorption of the photons. Thus, at a fixed $B$ field,
the peak positions in MO absorption spectrum relate to the energy spacings
between occupied and unoccupied LLs. By varying the magnetic field, the LL
energies should change (see Eq. \eqref{egvalue}) so that the peak positions can change
in the MO absorption spectrum (see Fig. \ref{fig10}). Meanwhile, the Fermi energy
also changes with the $B$ field. It is a fact that the LL occupancy in an
electron gas decreases with increasing magnetic field. As a result, when
the Fermi energy crosses over a LL by increasing $B$, some of the MO absorption
peaks can disappear owing to the closing down of the allowed optical transition
channels. Hence, by monitoring the disappearance of the MO absorption peaks
with altering the magnetic field, one can get the information about the Fermi
energy in an electron gas in different $B$ fields. Moreover, the main advantage
of the MO measurement is that it does not need to fabricate the Hall bar or
van de Pauw ohmic contacts on the sample. These contact electrodes are required
necessarily for measuring the longitudinal and transverse resistances in
conventional magneto-transport experiments. Therefore, the MO experiment is
a powerful and practical technique in studying and characterizing the electronic
materials and devices. In fact, in recent years the MO absorption spectra have
been utilized to probe and manipulate the spin and valley
polarized electrons and/or holes in ML-TMD systems \cite{Wang16}.

\section{Conclusions}
\label{sec:conclusions}
In this paper, we have presented a detailed theoretical study on Landau level
structure and longitudinal magneto-optical conductivity of ML-TMDs in the presence
of the proximity-induced exchange interaction and external electrical field.
By using the RPA dielectric function for a 2DEG in the presence of quantizing
magnetic field, the magneto-optical conductivity has evaluated through the
Kubo formula. We have examined the dependence of the MO conductivity or absorption
spectrum upon the proximity-induced exchange interaction, carrier density, external
electrical filed, and magnetic field in $n$- and $p$-type ML-TMDs. The main conclusions
obtained from this study can be summarized as follows.

The proximity-induced exchange interaction can break the spin and valley
degeneracies in the presence of a magnetic field which can introduce the
spin and valley polarizations. The spin and valley dependent electronic
transitions can result in two MO absorption peaks in THz regime for the
transitions within the conduction or valence LLs. A series of MO absorption
peaks can be observed in visible regime, which are attributed from electronic
transitions from valence LLs to conduction LLs. In addition, the proximity-induced
exchange interaction, carrier density, electrical field, and magnetic field
can effectively tune the positions of magneto-optical absorption peaks and
modulate the shape of the MO absorption spectrum. Experimentally, the MO
absorption of ML-TMDs in visible, infrared and THz ranges can be measured
by the techniques such as MO Fourier spectroscopy \cite{Delhomme19} and MO
THz time-domain spectroscopy \cite{Mei18,Wen21}. Therefore we hope the
theoretically findings shown and discussed in this paper can be verified
experimentally. It would be promising and interesting to detect the spin
and valley polarized electrons or holes through the characteristic magneto-optical
absorption peak. The obtained results in this study indicate that ML-TMDs
are advanced 2D electronic materials for realizing the magneto-optic, spintronic
and valleytronic devices which can work in visible and THz bandwidths. We
believe that the results obtained from this work can help us to gain an
in-depth understanding of the MO properties of ML-TMDs materials.

\section*{ACKNOWLEDGMENTS}
This work was supported by the National Natural Science foundation
of China (NSFC) (Grants No. 12004331, No. U2230122, No. U2067207,
and No. 12364009), Shenzhen Science and Technology Program (Grant
No. KQTD20190929173954826), and by Yunnan Fundamental Research
Projects (Grant No. 202301AT070120). Y.M.X was supported through the
Xingdian Talent Plans for Young Talents of Yunnan Province (Grant
No. XDYC-QNRC-2022-0492).

\end{document}